\documentclass[notitlepage,aps,prx,reprint,twocolumn,longbibliography,superscriptaddress]{revtex4-2}
\usepackage{graphicx}
\usepackage{amsmath}
\usepackage{amssymb}
\usepackage{comment}
\usepackage[colorlinks, allcolors=blue]{hyperref}
\usepackage[all]{hypcap}
\usepackage[mathlines]{lineno}
\usepackage{braket}
\usepackage{wrapfig}
\usepackage{lipsum}
\usepackage{ulem}

\newcommand{\pref}[2]{\hyperref[#1]{\ref{#1}(#2)}}
\newcommand{\preff}[2]{\hyperref[#1]{\ref{#1 b}#2}}
\newcommand{\eqpref}[1]{\hyperref[#1]{(\ref{#1})}}

\newcommand{\squig}{{\raise.17ex\hbox{$\scriptstyle\sim$}}}

\begin{document}
\title{Synthetic Mechanical Lattices with Synthetic Interactions}
\author{Ritika Anandwade}
\thanks{These authors contributed equally to this work.}
\author{Yaashnaa Singhal}
\thanks{These authors contributed equally to this work.}
\author{Sai Naga Manoj Paladugu}
\thanks{These authors contributed equally to this work.}
\affiliation{Department of Physics, University of Illinois at Urbana-Champaign, Urbana, IL 61801-3080, USA}
\author{Enrico Martello}
\thanks{These authors contributed equally to this work.}
\affiliation{School of Physics and Astronomy, University of Birmingham, Edgbaston, Birmingham B15 2TT, United Kingdom}
\author{Michael Castle}
\affiliation{Department of Physics, University of Illinois at Urbana-Champaign, Urbana, IL 61801-3080, USA}
\author{Shraddha Agrawal}
\affiliation{Department of Physics, University of Illinois at Urbana-Champaign, Urbana, IL 61801-3080, USA}
\author{Ellen Carlson}
\affiliation{Department of Physics and Astronomy, Haverford College, 370 Lancaster Ave, Haverford, PA 19041-1392, USA}
\author{Cait Battle-McDonald}
\affiliation{Department of Physics, Smith College, Northampton, MA 01063, USA}
\author{Tomoki Ozawa}
\email{tomoki.ozawa.d8@tohoku.ac.jp}
\affiliation{Advanced Institute for Materials Research (WPI-AIMR), Tohoku University, Sendai 980-8577, Japan}
\author{Hannah M. Price}
\email{H.Price.2@bham.ac.uk}
\affiliation{School of Physics and Astronomy, University of Birmingham, Edgbaston, Birmingham B15 2TT, United Kingdom}
\author{Bryce Gadway}
\email{bgadway@illinois.edu}
\affiliation{Department of Physics, University of Illinois at Urbana-Champaign, Urbana, IL 61801-3080, USA}
\date{\today}

\begin{abstract}
Metamaterials based on mechanical elements have been developed over the past decade as a powerful platform for exploring analogs of electron transport in exotic regimes that are hard to produce in real materials. In addition to enabling new physics explorations, such developments promise to advance the control over acoustic and mechanical metamaterials, and consequently to enable new capabilities for controlling the transport of sound and energy.
Here, we demonstrate the building blocks of highly tunable mechanical metamaterials based on real-time measurement and feedback of modular mechanical elements. We experimentally engineer synthetic lattice Hamiltonians describing the transport of mechanical energy (phonons) in our mechanical system, with control over local site energies and loss and gain as well as control over the complex hopping between oscillators, including a natural extension to non-reciprocal hopping. Beyond linear terms, we experimentally demonstrate how this measurement-based feedback approach opens the window to independently introducing nonlinear interaction terms.
Looking forward, synthetic mechanical lattices open the door to exploring phenomena related to topology, non-Hermiticity, and nonlinear dynamics in non-standard geometries, higher dimensions, and with novel multi-body interactions.
\end{abstract}
\maketitle

Networks of coupled harmonic oscillators have long served as a foundational model for understanding thermal transport in solids~\cite{Debye}, and over the past decade have additionally become a powerful theoretical and experimental platform for exploring topology~\cite{Huber2016}
and its connections to mechanical structures~\cite{TM-Kane,TM-LubenskyRev,Ma2019}.
While experiments based on physically coupled oscillators offer powerful capabilities for the realization of artificial materials and the visualization of novel transport phenomena therein~\cite{HuberPendulum,GyroscopeMetamaterials,fidgetspinners,TM-Prodan,TM-Huber-classes,TM-Huber-WeakTI,TM-Huber-SOC,Salerno_2017NJP,Camelia-DynPumping,Camelia-FlatBands,TM-Bahl-Pumping,Prodan-passive,Serra-Garcia2018}, such physical coupling terms present natural limitations on the Hamiltonians that may be directly engineered.
For example, Newton's third law dictates that the direct hopping terms should obey reciprocity, with forward and backward tunneling pathways having equal amplitudes. The position-dependence of spring forces further implies the restriction to realizing only time-reversal invariant hopping Hamiltonians.

While a number of clever approaches have been proposed~\cite{Salerno-alpha,Salerno_2014_epl,Wang_2015,Mitchell2018}
and implemented~\cite{HuberPendulum,GyroscopeMetamaterials} to circumvent such limitations while maintaining physical coupling between oscillators, one may seek alternative approaches that avoid direct physical connections altogether.
In the context of classical electrical or mechanical metamaterials, where mode occupations are on order of the Avogadro number, one can naturally think about utilizing measurements of the oscillators' properties - \textit{e.g.}, center-of-mass positions and momenta - as a resource for Hamiltonian engineering, with little concern for the disturbance of the natural system dynamics.
Indeed, the natural suitability of classical metamaterials for measurement-based feedback has in recent years led to proposals for the realization of designer non-Newtonian systems~\cite{Ilan-prop}, and even first demonstrations of the engineering of non-reciprocity in robotic mechanical metamaterials of physically coupled rotors~\cite{Brandenbourger2019,Coulais-PNAS}.
Here, through the measurement of and feedback on a system of otherwise physically disconnected mechanical oscillators, we demonstrate a general approach to engineering nonlinear synthetic lattice Hamiltonians.
We experimentally demonstrate the engineering of complex and non-reciprocal hopping terms, complex local site energies, and synthetic quartic nonlinearities, exploring the use of feedback-based control to drive both $\mathcal{PT}$-symmetry breaking and Josephson self-trapping phase transitions in a synthetic double-well.
The extension to larger, many-site arrays of synthetically coupled oscillators, incorporating even more exotic synthetic nonlinearities, will enable explorations of novel lattice Hamiltonians with tailored mean-field interactions.

This paper is organized as follows. Section~\ref{desc} describes our experimental system and presents an informal discussion of our feedback-based approach to Hamiltonian engineering. In Sec.~\ref{theory}, we provide the formal theoretical framework underlying this approach to engineering effective tight-binding models for mechanical oscillators synthetically coupled by measurement-based feedback. In Sec.~\ref{examps}, we provide several examples for the engineering of specific Hamiltonian terms, such as local site energies (real and imaginary) and inter-site hopping terms (complex and non-reciprocal), as well as nonlinear interaction terms. For each example, we provide a theoretical derivation for the required feedback forces, as well as an experimental demonstration of the implementation. We summarize our results in Sec.~\ref{conc}.

\begin{figure*}[t]
	\includegraphics[width=1.98\columnwidth]{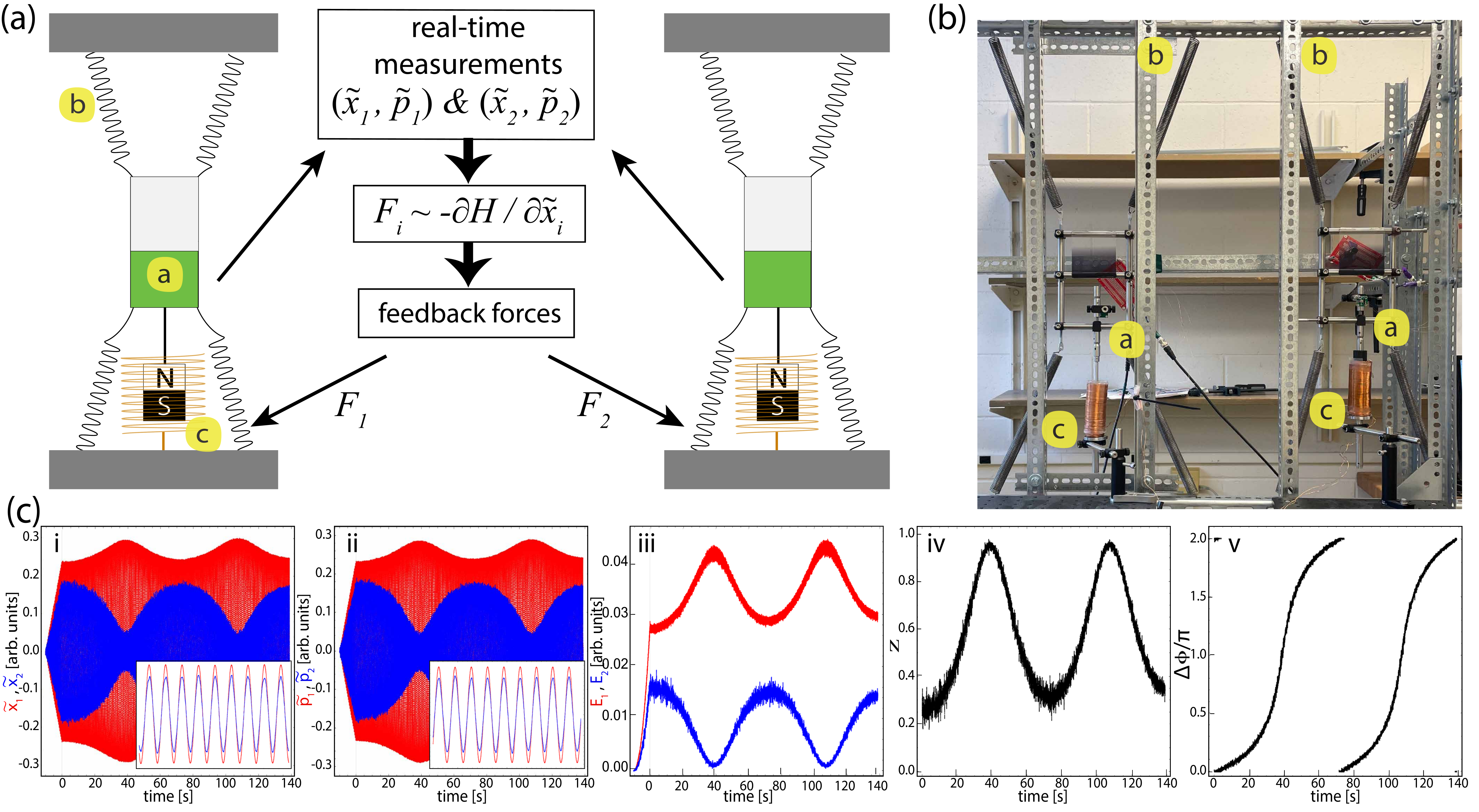}
	\caption{\label{FIG:fig1}
		\textbf{Modular mechanical oscillators synthetically coupled by measurement-based feedback.}
		\textbf{(a)}~A cartoon depiction of the implemented mechanical oscillators, which feature embedded accelerometers (marked $a$) for the real-time measurement of proxies for position ($\tilde{x}$) and momentum ($\tilde{p}$), a set of four springs (one marked $b$), and a dipole magnet embedded in a gradient solenoid for the application of forces (marked $c$).
		Real-time feedback forces $F_i$, which depend on the real-time measurements $\tilde{x}_i$ and $\tilde{p}_i$, are used to implement an effective tight-binding Hamiltonian $H$.
		\textbf{(b)}~A photograph of the large-scale prototype used to implement the two-site synthetically-coupled mechanical lattice depicted in (a), with letters denoting the same elements.
		\textbf{(c)}~Acquired experimental data and processed observables for a two-site system with synthetic hopping and synthetic nonlinearity. 
		Panel \textbf{i}: Experimental measurements of $\tilde{x}_1$ (red) and $\tilde{x}_2$ (blue), with inset showing short-time dynamics over several oscillator periods from $t = 0$ to 3~s.
		Panel \textbf{ii}: Similar plots for the corresponding $\tilde{p}_i$ measurements.
		Panel \textbf{iii}: Constructed proxy for the local mechanical energy, $E_i \propto \tilde{x}_i^2 + \tilde{p}_i^2$, for two coupled masses with synthetic nonlinearity.
		Panel \textbf{iv}: Dynamics of the normalized population imbalance $z = (E_1 - E_2) / (E_1 + E_2)$.
		Panel \textbf{v}: Dynamics of the relative oscillator phase $\Delta \phi = \phi_1 - \phi_2$, where the local oscillator phase is reconstructed from the $\tilde{x}$ and $\tilde{p}$ measurements as $\phi_i = \textrm{arg}(\tilde{x}_i + i \tilde{p}_i)$.
    	}
\end{figure*}

\section{Description of the system}
\label{desc}

As depicted in Fig.~\pref{FIG:fig1}{a,b}, our prototype for a ``lattice of synthetically coupled oscillators'' consists of modular, large-scale (kg-scale mass) mechanical oscillators. In the absence of applied feedback forces, these oscillators are characterized by nearly identical natural oscillation frequencies $f_0\sim$~3.05~Hz and quality factors $Q\sim$~1000.

An analog accelerometer (EVAL-ADXL203) is fixed to each oscillator, and we acquire real-time measurements of acceleration $a(t)$ by sending the signals to a common computer. By taking the numerical derivative of the acquired signal, we additionally acquire real-time measurements of the oscillators' jerk $j(t) \equiv \partial a(t) / \partial t$. As the signals come from harmonic oscillators with roughly constant frequencies, we can associate the measured acceleration and jerk signals as proxies for the oscillator position $x(t)$ and momentum $p(t)$ signals, respectively (as the sets of variables $\{ a, x\}$ and $\{j, p\}$ have proportional relationships, as we explain in more detail later). We hereafter refer to the input signals as position $\tilde{x}$ and momentum $\tilde{p}$.
In experiment, we normalize the $\tilde{x}$ and $\tilde{p}$ signals to the same dimensionless amplitude, reflecting the equipartition of kinetic and potential energy.

Our ``synthetic mechanical lattice'' approach implements an effective Hamiltonian $\mathcal{H} = H + H_0$ in the oscillator array, where the portion $H$ can be considered as a perturbation to the bare Hamiltonian $H_0$ of the uncoupled, identical oscillators.
The modified part of the Hamiltonian, $H$ (the terms of which have frequency scales $\ll f_0$), describes the transport of mechanical energy (phonons) between the oscillators, small shifts to the oscillator frequencies, and any engineered nonlinearities.
Roughly speaking, to implement
$H$ we apply individual feedback forces to the oscillators that reflect the relationship $F_i \sim -\partial H / \partial \tilde{x}_i$ (\textit{cf.} Fig.~\pref{FIG:fig1}{a}).
These forces can in principle have almost any dependence on the positions $\tilde{x}_i$ and momenta $\tilde{p}_i$, including higher powers thereof, opening up new possibilities for Hamiltonian engineering~\footnote{Here we consider only instantaneous dependencies, but this approach also allows for time-retarded interactions}. To note, we will develop the framework for this feedback-based control more formally in Sec.~\ref{theory}.

We implement these feedback forces
magnetically, avoiding any added mechanical contacts. Each oscillator has a dipole magnet attached to a central, cylindrical shaft. The dipole magnet is embedded in a wound coil (gradient solenoid~\cite{SolenoidDesign}, with a higher winding density at its base than at its top). We control the current (between 0 and 2~A) in the coil, which produces an axial magnetic field gradient that in turn creates a force on the oscillator. While there is a fixed direction of current flow, we operate with a nominal offset gradient and control the variations about this offset, thus achieving an effective bi-directional (positive and negative along the axial direction) control of forces.

Figure~\pref{FIG:fig1}{c} displays the typical real-time measurements acquired for one example experiment, which explores a self-trapped mode in a nonlinear double-well (discussed further later on, in the context of Fig.~\ref{FIG:fig6}).
The measured $\tilde{x}$ (panel \textbf{i}) and $\tilde{p}$ (panel \textbf{ii}) signals for oscillators 1 (red) and 2 (blue) are shown in panels \textbf{i} and \textbf{ii}, including zoomed in views over several seconds showing the intra-envelope dynamics.
From these primary measurements, we construct a proxy for the local mechanical energy $E_i = (\tilde{x}_i^2 + \tilde{p}_i^2)/2$, as shown in panel \textbf{iii}. The local mechanical energy plays a role analogous to the local particle probability density $|\psi_i|^2$ of a wave function $\psi$ under the evolution of the implemented tight-binding Hamiltonian. 
One may also extract the local phase $\phi_i = \textrm{arg}(\tilde{x}_i + i\tilde{p}_i)$ at each oscillator, associated with $\textrm{arg}(\psi_i)$ of the corresponding evolving wave function.
To note, the initial linear and quadratic rise of the signals in panels \textbf{i}/\textbf{ii} and \textbf{iii}, respectively, relate to an initial preparation step of 10~s during which a sinusoidal force prepares the respective energy and phases of the two oscillators.
In panels \textbf{iv} and \textbf{v}, we plot further derived experimental quantities relevant to the dynamics in this case of a tunnel-coupled double-well with nonlinear interactions.
Panel \textbf{iv} depicts the normalized energy imbalance $z = (E_1 - E_2) / (E_1 + E_2)$ and panel \textbf{v} depicts the relative oscillator phase $\Delta \phi = \phi_1 - \phi_2$. The trajectories of $z$ and $\Delta \phi$ reflect a self-trapped mode with a trapped population imbalance but a running relative phase.


\section{Mapping and theory background}
\label{theory}

The theoretical basis for our synthetic mechanical metamaterial is a mapping which can be made in certain limits from Newton's equations of motion to the Heisenberg equations for a tight-binding quantum Hamiltonian. Such an approach has previously been used, for example, to propose how to simulate a Peierls hopping phase and an effective Harper-Hofstadter model with time-modulated classical coupled harmonic oscillators~\cite{Salerno-alpha,Salerno_2014_epl}.
In this section, we show how this approach can be applied to implement a wide variety of Hamiltonian terms by subjecting individual and pairs of classical oscillators to weak feedback.

Before entering into details of the mapping, we start by considering the equations of motion for a pair of uncoupled and identical harmonic oscillators:
\begin{eqnarray}
m \dot{x}_i(t) = p_i(t) , \qquad   \dot{p}_i(t) = - m \omega^2 x_i(t), 
\end{eqnarray}
where $x_i(t)$ and $p_i(t)$ are position and momentum of an oscillator at time $t$ with the index $i=1,2$ running over the two oscillators, $\omega\!=\! \omega_1 \!=\! \omega_2$ is the angular oscillation frequency and $m \!=\!m_1\!=\! m_2$ is the mass. The dot over variables denote the first derivative in time. Below, we often suppress the explicit time-dependence, $(t)$, when it is obvious. For convenience of measurement, it is more natural for us to work with the acceleration, $a_i(t)$, and the jerk, $j_i(t) \equiv \dot{a}_i(t)$, instead of the position and momentum. However, for a harmonic oscillator, these are simply related as 
\begin{align}
    a_i(t) &= - \omega^2 x_i (t), & j_i(t) = -\frac{\omega^2}{m} p_i(t)
\end{align}
and their equations of motion are
\begin{align}
    \dot{a}_i(t) &= j_i(t), & \dot{j}_i(t) &= -\omega^2 a_i(t)
\end{align}
suggesting the acceleration as a proxy for the position, and the jerk as a proxy for the momentum. To make this concept clearer in what follows, we shall introduce the notation $X_i \equiv  a_i$ and $P_i \equiv  j_i$, so that the equations of motion  become
\begin{eqnarray}
\dot{X}_i = P_i , \qquad \dot{P}_i = - \omega^2 X_i . 
\end{eqnarray}

To couple the two oscillators and simulate different effects, we now add feedback to the system such that the equations of motion become
\begin{eqnarray}
\dot{X}_i = P_i , \qquad \dot{P}_i = - \omega^2 X_i + F_i  . 
\end{eqnarray}
where $F_i$ is a function of $(X_1, X_2, P_1, P_2)$. In analogy with the real momentum, this feedback acts as a ``force'' on the oscillator. Here we present a general recipe, while below, we shall discuss specific examples of $F_i$ that we consider to map to different Hamiltonians.
As stated above, the experiment works with normalized effective variables for position and momentum. However, for concreteness, we retain the dimensionality of $X_i$ and $P_i$ and include relevant angular frequency $\omega$ terms
below.

To map the above equations to Heisenberg equations, we introduce the classical complex variables~\cite{Salerno-alpha,Salerno_2014_epl}:
\begin{eqnarray}
\alpha_i \equiv \sqrt{\frac{\omega}{2}} X_i  + i \sqrt{\frac{1}{2 \omega}} P_i 
\end{eqnarray}
in analogy with the annihilation operator of the quantum harmonic oscillator. It can be straightforwardly shown from this that $|\alpha_i|^2$ then scales with the instantaneous oscillation energy of a given mass.  
From this, it follows that we can re-express the acceleration and jerk as
\begin{eqnarray}
X_i= \sqrt{\frac{1}{2 \omega}} (\alpha_i + \alpha_i^*), \qquad P_i =- i \sqrt{\frac{\omega}{2}} ( \alpha_i - \alpha_i^*) \label{eq:XP}
\end{eqnarray}
and hence the equations of motion are
\begin{eqnarray}
\dot{\alpha}_i = - i \omega \alpha_i + \frac{i}{\sqrt{2 \omega}} {F}_i , \label{eq:dynamics}
\end{eqnarray}
where the feedback term, $F_i$, 
should also be re-expressed as a function of $(\alpha_1, \alpha_1^*, \alpha_2, \alpha_2^*)$. Note that the complex conjugate of this equation describes the time-evolution of the conjugate variables, $\alpha_i^*$. 

In the absence of the feedback, 
it can be seen from Eq.~\ref{eq:dynamics} that the time-dependence of the complex amplitudes is given by $\alpha_i (t) \propto e^{-i \omega t}$ (and similarly $\alpha_i^* (t) \propto e^{i \omega t}$) as expected for harmonic oscillators. 
Including the feedback naturally modifies the dynamics. However, provided that $\omega$ remains the largest frequency-scale in the problem and that the
feedback is sufficiently weak, these dynamical changes will be slow and small compared to the natural oscillations. In this limit, we can assume that the complex amplitudes' fastest time-dependence is still given by $\alpha_i (t) \propto e^{-i \omega t}$~\cite{Salerno-alpha,Salerno_2014_epl}, or in other words,  that the $\alpha_i$ variables ``rotate'' with a frequency $\approx \omega$ (while the conjugate variables $\alpha_i^*$ ``rotate" with $\approx - \omega$). 

Working in this high-frequency limit allows us to apply the ``rotating-wave approximation'' (RWA) to simplify Eq.~\ref{eq:dynamics}~\cite{Salerno-alpha,Salerno_2014_epl}. The RWA is an approach well-known from quantum optics, in which only so-called ``co-rotating terms'' with a frequency $\approx \omega$ are kept in the dynamics. To physically understand the RWA, we can imagine transforming Eq.~\ref{eq:dynamics} into a ``co-rotating frame'' at the natural frequency $\omega$. In this frame, a term $\propto \alpha_i$, for example, varies relatively slowly (due to feedback), while a term $\propto \alpha_i^*$ oscillates rapidly at a frequency $\approx - 2 \omega$. Terms like the former are ``co-rotating'' while the latter are  ``counter-rotating'' as they will rapidly average to zero over the timescale for the slow dynamics. In the limit that $\omega \rightarrow \infty$, only ``co-rotating terms'' with a frequency $\approx \omega$ remain important, justifying the RWA assumption that all other terms can be dropped. 

Under the condition of high natural frequency and weak feedback, we can therefore re-write Eq.~\ref{eq:dynamics} as
\begin{eqnarray}
\dot{\alpha}_i = - i \omega  \alpha_i + \sum_{j
}\frac{i}{\sqrt{2 \omega}} {F}_{i j}^{\text{RWA}}  \alpha_j \ ,
\label{eq:RWA}
\end{eqnarray}
where $F_i \rightarrow \sum_{j}{F}_{ij}^{\text{RWA}} \alpha_j $ is the  RWA, \textit{i.e.}, where we only keep suitable co-rotating terms in the feedback applied to oscillator $i$. Note that such terms must contain at least one factor of either $\alpha_1$ or $\alpha_2$ (as both oscillators have natural frequency $\omega$), and so we have explicitly factored out $\alpha_j$ in this expression for convenience. In general, ${F}_i^{\text{RWA}} $ can still be any suitable function of $(\alpha_1, \alpha_1^*, \alpha_2, \alpha_2^*)$, as discussed further below, and so can encode a wide-range of physical effects. Explicit examples are given in the following sections to further illustrate this approximation. 

The equation of motion (Eq.~\ref{eq:RWA}) can be regarded as the Heisenberg equation of motion derived from a quantum-mechanical Hamiltonian (with $\hbar$ suppressed)
\begin{align}
    \mathcal{H}
    =
    \sum_i \omega \hat{\alpha}_i^\dagger \hat{\alpha}_i - \frac{1}{\sqrt{2\omega}}\sum_{i,j} \frac{1}{n_i+1} \hat{\alpha}_i^\dagger \hat{F}_{ij}^\mathrm{RWA} \hat{\alpha}_j, \label{eq:H}
\end{align}
where $\hat{\alpha}_i^\dagger$ and $\hat{\alpha}_i$ are creation and annihilation operators for site $i$ obeying bosonic commutation relations.
The operator $\hat{F}_{ij}^\mathrm{RWA}$ is an operator obtained by replacing $(\alpha_1, \alpha_1^*, \alpha_2, \alpha_2^*)$ in ${F}_{ij}^{\text{RWA}}$ by $(\hat{\alpha}_1, \hat{\alpha}_1^\dagger, \hat{\alpha}_2, \hat{\alpha}_2^\dagger)$ and moving all the creation operators to the left of the annihilation operators. The factor of $n_i$ is the number of the operators $\alpha_i^\dagger$ appearing in ${F}_{ij}^{\text{RWA}}$.
With the identification $\alpha_i = \langle \hat{\alpha}_i\rangle$ and $\alpha_i^* = \langle \hat{\alpha}_i^\dagger\rangle$, where $\langle \cdot \rangle$ is the quantum mechanical average of the operator with respect to the initial state, the Heisenberg equation of motion for Eq.~\ref{eq:H} reduces exactly to Eq.~\ref{eq:RWA} when $\hat{F}_{ij}^\mathrm{RWA}$ does not itself contain operators. When $\hat{F}_{ij}^\mathrm{RWA}$ does contains operators, this will correspond to inter-particle interactions in the quantum-mechanical language, as discussed further below. In such cases, assuming that the energy in each oscillator is large enough so that we are in the classical limit, we can approximate $\langle \hat{F}_{ij}^\mathrm{RWA} \hat{\alpha}_j \rangle \approx F_{ij}^\mathrm{RWA} \alpha_j$ such that interaction terms also reduce to the corresponding terms in Eq.~\ref{eq:RWA}.

Hence, the dynamics of our synthetic mechanical metamaterial can be used to simulate the above general class of quantum mechanical tight-binding Hamiltonians. Note also that although we focus experimentally in this paper on up to two oscillators, the above equations are valid for $n$ identical oscillators, in which case $\mathcal{H}$ will describe an $n$-site Hamiltonian. The generalization to non-identical oscillators is also straightforward, as discussed in Ref.~\cite{Salerno-alpha}, provided that all the natural oscillator frequencies are large enough that the RWA can be applied. 

Having reviewed the above general theoretical recipe, we shall now give explicit examples for using feedback to engineer specific types of desired terms in the tight-binding Hamiltonian, both in theory and experiment.

\section{Hamiltonian control examples}
\label{examps}

\subsection{Control of Local Site Energy Shifts}

As a first example of this theoretical approach, we show how adding weak feedback corresponding to $F_i \propto X_i$ leads to local site energy shifts in the above Hamiltonian mapping. Physically, such a term corresponds in experiment to applying a local feedback which is proportional to the measured acceleration of the given oscillator.

Theoretically, we start by writing the feedback term as
\begin{eqnarray}
F_i = A_{i,i} X_i = \frac{A_{i,i}}{\sqrt{2 \omega}} ( \alpha_i + \alpha_i^*), \label{eq:xfeedback}
\end{eqnarray}
where $A_{i,i}$ is the (weak) amplitude for the feedback applied to oscillator $i$, which depends on the measurements of $X_i$. (Note that Einstein's index summation convention is not applied here).

\begin{figure}[t!]
	\includegraphics[width=1\columnwidth]{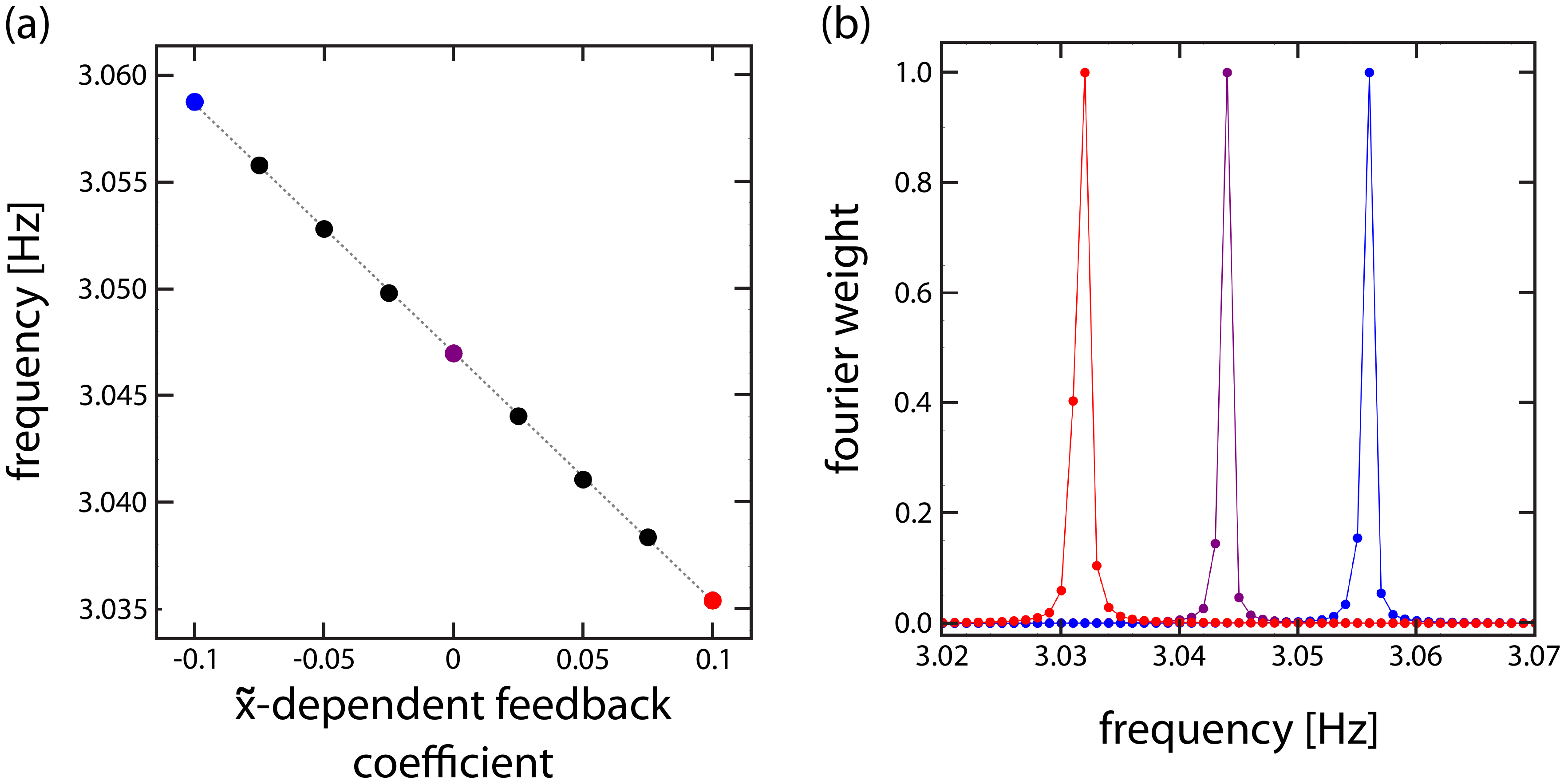}
	\centering
	\caption{\label{FIG:fig2}
		\textbf{Local control of site energy shifts by position-dependent feedback.}
        \textbf{(a)}~Oscillation frequency of an individual oscillator (determined by Fourier analysis of the experimental oscillator dynamics), 
        as a function of the coefficient for the position-dependent feedback force.
        Error bars (smaller than the data points) represent the standard error of the mean of the fit used to determine the points.
        \textbf{(b)}~Scaled power spectra of the Fourier-transformed oscillator position dynamics for three different values of the $\tilde{x}$ feedback coefficient, with colors relating to those of the points in (a).
	}
\end{figure}

Adding a feedback like this means that one of our equations of motion becomes $\dot{P}_i = - (\omega^2 - A_{i,i} ) X_i$, which is equivalent to having a new effective natural frequency 
$\tilde{\omega}_i = \sqrt{\omega^2 - A_{i,i}}$.
Assuming weak amplitudes for the driving force and taking a Taylor expansion of this expression, we obtain
$\tilde{\omega}_i \simeq \omega - A_{i,i}/2\omega$.

Alternatively, this result can be obtained by substituting the applied feedback form into Eq.~\ref{eq:dynamics}, leading to
\begin{eqnarray}
\dot{\alpha}_i = - i \omega \alpha_i + i \frac{A_{i,i}}{{2 \omega}} ( \alpha_i + \alpha_i^*) .  
\end{eqnarray}
Applying the RWA means that we neglect the counter-rotating $\alpha_i^*$ term such that
\begin{eqnarray}
 {F}_{i j}^{\text{RWA}} \alpha_j  = \begin{cases}
\frac{A_{i ,i}}{\sqrt{2 \omega}} \alpha_i 
& \text{for } i=j\\
0 & \text{for } i\neq j
\end{cases}
\end{eqnarray}
The Hamiltonian [Eq.~\ref{eq:H}] then follows directly as
\begin{eqnarray}
\mathcal{H} = \sum_{i}  ( \omega  + \Delta_i) \hat{\alpha}_i^\dagger \hat{\alpha}_i  ,
\end{eqnarray}
where $\Delta_i = -A_{i,i}/ 2 \omega$.
In terms of the tight-binding Hamiltonian, this is interpreted as adding a tunable site energy shift
$\Delta_i$
to
site $i$ of the
lattice, as desired. 
Physically, this corresponds to a local shift of the
corresponding oscillator frequency by an amount $\Delta_i$. 

We now demonstrate experimentally this control of local site energies by position-dependent self-feedback.
Figure~\ref{FIG:fig2} depicts the control of the oscillator frequency, corresponding to site energy shifts in the equivalent tight-binding model, through the application of $\tilde{x}$-dependent self-feedback. In Fig.~\pref{FIG:fig2}{a}, we plot the experimentally measured oscillator frequency as a function of the applied coefficient of self-feedback. We find a linear relationship between the measured frequency and the applied feedback, consistent with the fact that we are safely in the limit of weak feedback, with the modifications to the bare oscillator frequency at the level of $\lesssim$~0.5$\%$.

These oscillation frequencies are determined by first depositing mechanical energy into the oscillator over 10~s (by application of a strong oscillating force) and then allowing the oscillator to ring down freely over 100~s. We then perform a numerical Fourier transform of the position data and fit the resulting spectrum to a Lorentzian line shape, extracting the center frequency. Figure~\pref{FIG:fig2}{b} shows several such experimental line shape curves, with colors corresponding to the data points of the extracted center frequencies in Fig.~\pref{FIG:fig2}{a}.

\begin{figure}[b!]
	\includegraphics[width=1\columnwidth]{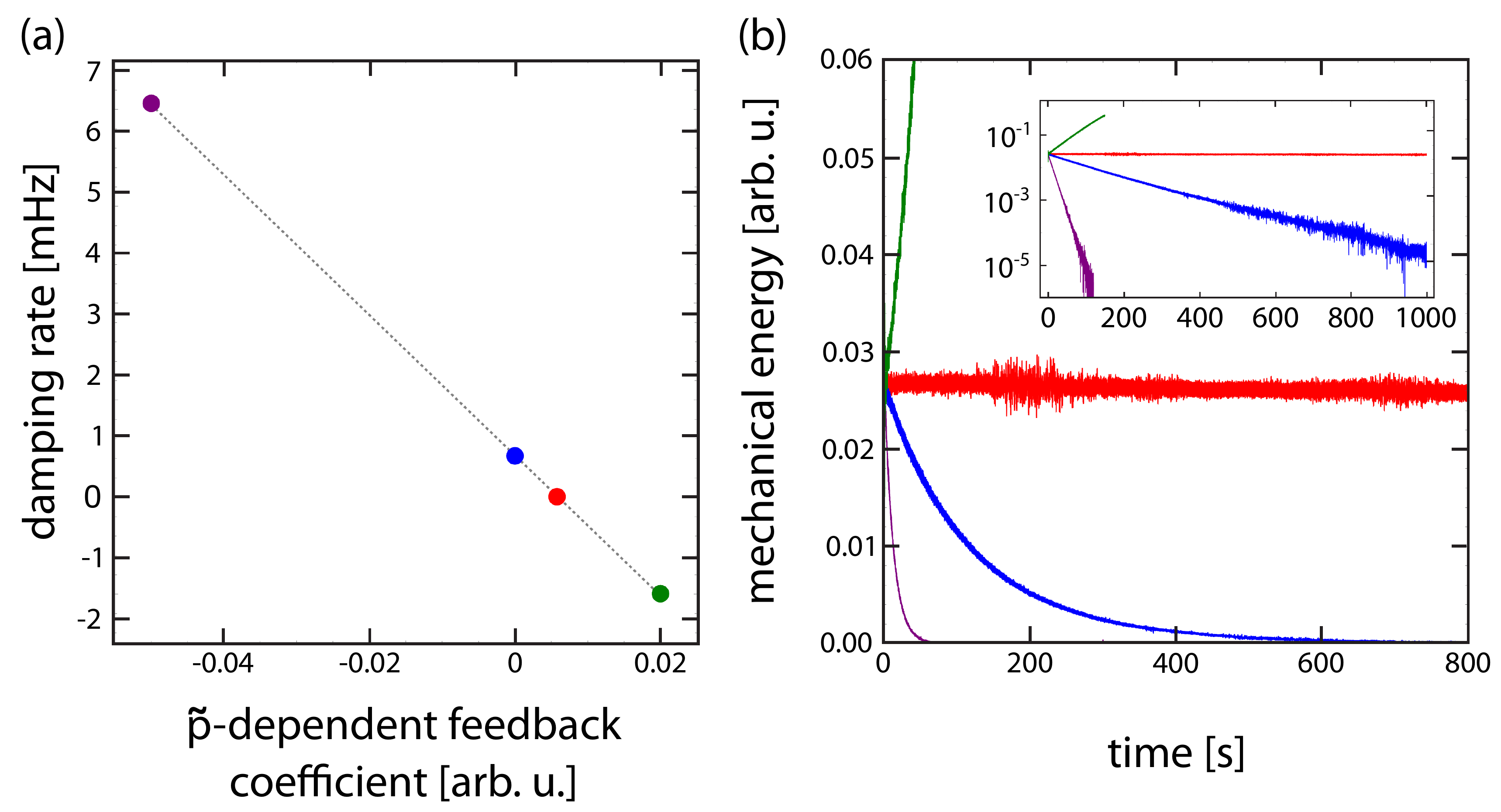}
	\centering
	\caption{\label{FIG:fig2-loss}
		\textbf{Local control of loss and gain by momentum-dependent feedback.}
        \textbf{(a)}~Damping rate of an individual oscillator as a function of the applied coefficient of momentum-dependent feedback. The damping rate $\gamma$ for each point is determined by fitting experimental data sets of the mechanical energy dynamics to an exponential decay $\propto e^{-2(2\pi\gamma t)}$, with negative $\gamma$ values relating to gain.
        Error bars (smaller than the data points) represent the standard error of the mean of the fit.
        \textbf{(b)}~Mechanical energy dynamics for several values of the applied $\tilde{p}$ feedback coefficient. Colors relate to those of the points in (a). Inset: semi-log plot of the same oscillator energy growth/decay curves.
	}
\end{figure}

\begin{figure*}[t]
	\includegraphics[width=1.85\columnwidth]{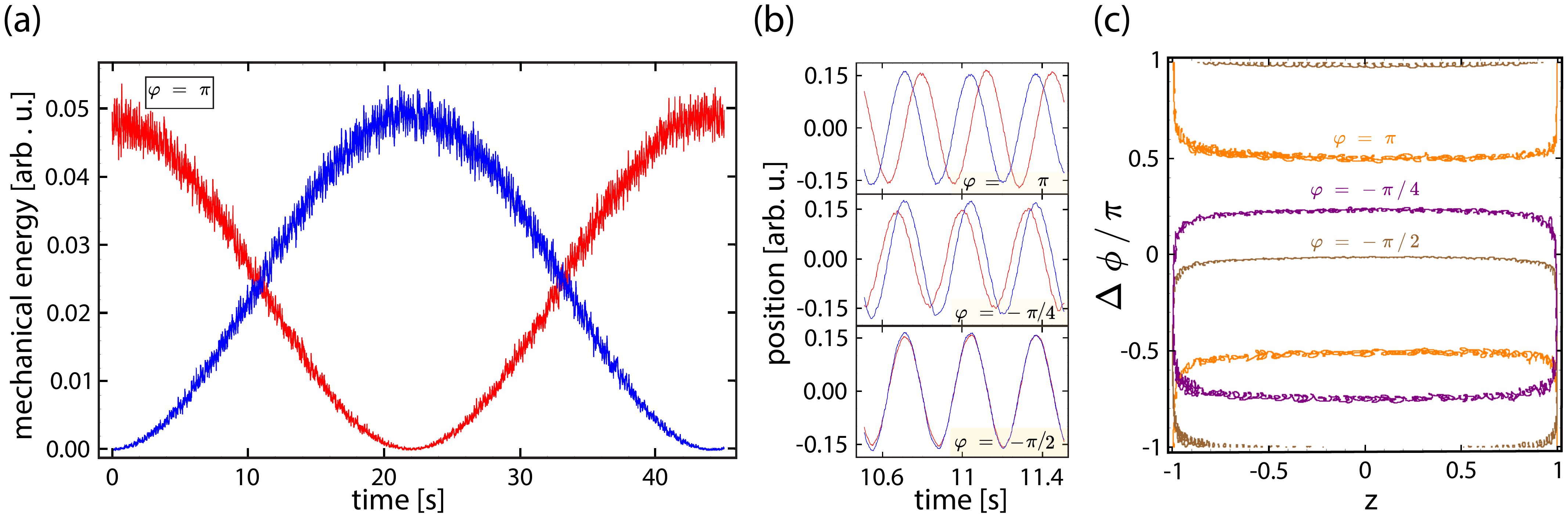}
	\centering
	\caption{\label{FIG:fig3}
		\textbf{Inter-oscillator hopping.}
        \textbf{(a)}~Transfer of mechanical energy between two synthetically coupled oscillators as a function of time for a tunnelling phase $\varphi = \pi$. The red and blue data curves relate to the experimentally measured mechanical energy dynamics of oscillators 1 and 2, respectively.
        \textbf{(b)}~Experimental oscillator position ($\tilde{x}$) dynamics over several oscillation periods, for the cases of tunneling phase values $\varphi = \pi$ (top), $-\pi/4$ (middle), and $-\pi/2$ (bottom).
        \textbf{(c)}~Phase space maps depicting the dynamical trajectories of the measured normalized mechanical energy imbalance $z$ and relative oscillator phase $\Delta \phi$ of the oscillator double-well. The trajectories related to the three tunneling phase values depicted in (b), for $\varphi = \pi$ (orange), $\varphi = -\pi/4$ (purple), and $\varphi = -\pi/2$ (brown). The trajectories reflect data sampled from one period of the mechanical energy dynamics ($\sim$50~s), with a low-pass filter applied to the plotted data.
}
\end{figure*}

\subsection{Local Control of Loss and Gain}

Similarly, local feedback can be used to simulate on-site loss and gain. For a tight-binding Hamiltonian, this would correspond to having
\begin{eqnarray}
\mathcal{H} = \sum_{i}  ( \omega  - i \gamma_i) \hat{\alpha}_i^\dagger \hat{\alpha}_i ,
\end{eqnarray}
where $\gamma_i$ is real and with $\gamma_i>0$ and $\gamma_i<0$ representing local loss and gain, respectively. From this, we can read-off that we require 
\begin{eqnarray}
{F}_{i j}^{\text{RWA}} \alpha_j  = \begin{cases}
i \gamma_i \sqrt{2 \omega} \alpha_i 
& \text{for } i=j\\
0 & \text{for } i\neq j
\end{cases}
\end{eqnarray}
It is straightforward to see from Eq.~\ref{eq:XP} that a suitable choice of feedback is
\begin{eqnarray}
F_i = B_{i,i} P_i = - i B_{i,i}\sqrt{\frac{\omega}{2}} ( \alpha_i - \alpha_i^*), 
\end{eqnarray}
where $\gamma_i = - B_{i,i}/2 $, as the unwanted $\alpha_i^*$ term in this expression will be neglected in the RWA. Here, $B_{i,i}$ is the (weak) amplitude for the feedback applied to oscillator $i$, which depends on the measurements of $P_i$. Physically, this corresponds in our setup to local feedback which is proportional to the jerk of the oscillator.

Our control of the local loss and gain terms of a single oscillator (site) is demonstrated in Fig.~\ref{FIG:fig2-loss}.
Figure~\pref{FIG:fig2-loss}{a} plots the measured rate of mechanical energy damping as a function of the applied coefficient of momentum-dependent self-feedback. We find a very linear relationship between the applied amplitude of feedback and the shift of the measured rate of damping/gain (with a nearly identical slope to that found in Fig.~\pref{FIG:fig2}{a}, owing to our normalization of the dimensionless $\tilde{x}$ and $\tilde{p}$ signals).

The damping rates in Fig.~\pref{FIG:fig2-loss}{a} are determined as follows. We first excite the oscillator (as in Fig.~\ref{FIG:fig2}) by sinusoidally driving it near resonance for 10~s, and then allow it to undergo free evolution under the applied feedback for up to 1000~s. We then compute a proxy of the mechanical energy stored in the oscillator, $E = (\tilde{x}^2 + \tilde{p}^2)/2$, as plotted for the four different feedback cases shown in Fig.~\pref{FIG:fig2-loss}{b}.
A simple exponential decay curve is fit to the dynamics of the mechanical energy, with negative decay values indicating an exponential gain. Starting from a relatively long decay time of $\sim 200$~s, corresponding to a natural quality factor of nearly 1000, we can introduce greatly enhanced loss or even strong gain through this momentum-dependent feedback. Importantly, for the investigation of unitary dynamics, we can also achieve an excellent cancellation of the loss/gain terms by the appropriate weak self-feedback (\textit{cf.} red curve in Fig.~\pref{FIG:fig2-loss}{b}).
We note that similar feedback for the enhancement of mechanical quality factors has previously been reported in Ref.~\cite{TM-Bahl-Pumping}.

\begin{figure*}[t]
	\includegraphics[width=1.92\columnwidth]{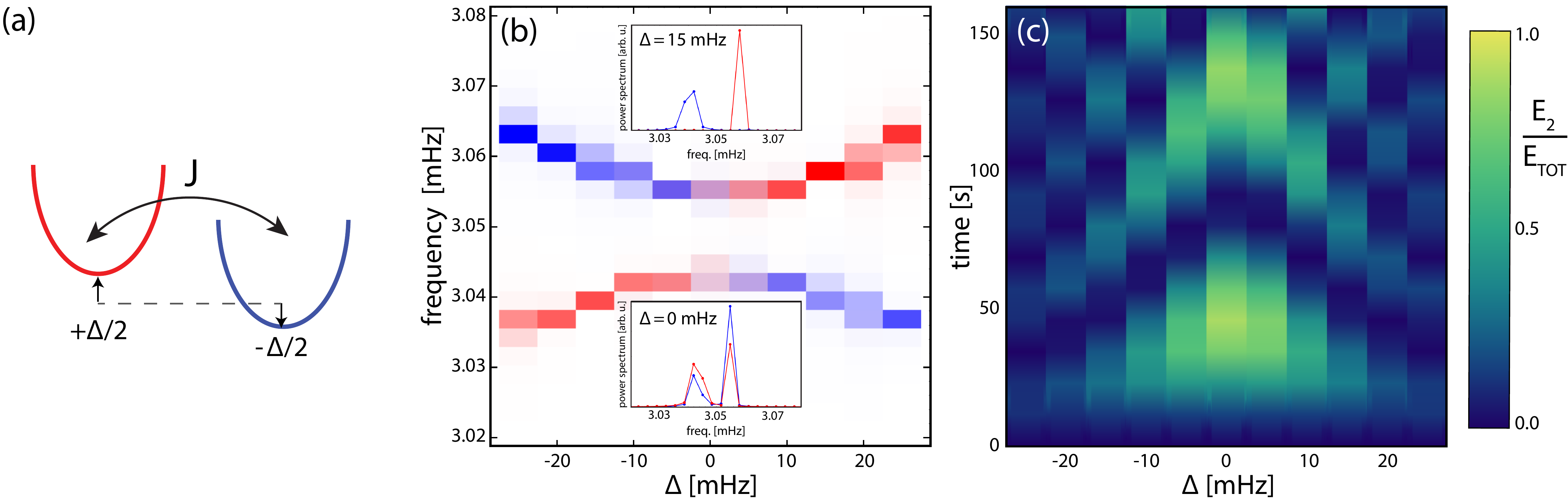}
	\centering
	\caption{\label{FIG:fig4}
		\textbf{Avoided crossing in a biased double-well.}
        \textbf{(a)}~A two-oscillator ``double-well'' with synthetic hopping $J$ and an inter-well bias $\Delta$.
        \textbf{(b)}~Frequency spectra of the two-oscillator system as a function of inter-well bias $\Delta$. Overlaid power spectra of the coupled oscillator dynamics upon initialization of oscillator 1 (red) and oscillator 2 (blue), as determined by a Fourier-transform of the experimental dynamics over 380~s. White regions relate to frequencies with no weight in either oscillator, purple relates to delocalized modes, and red (blue) regions relate modes with weight concentrated in oscillator 1 (2).
        Insets show the power spectra when initializing in oscillator 1 (red) and 2 (blue) for values of $\Delta = 15$~mHz (top) and $\Delta = 0$~mHz (bottom).
        \textbf{(c)}~Population dynamics as a function of $\Delta$. Initializing mechanical energy solely in oscillator 1, the plot shows the experimental dynamics of the mechanical energy appearing in oscillator 2, normalized to the total energy.
	}
\end{figure*}

\subsection{(Complex) Inter-oscillator Hopping}

So far, only self-feedback of the form $F_{ij}^{\text{RWA}}\alpha_j \propto \delta_{ij}$ has been taken into consideration and, as a consequence, only on-site phenomena have been investigated. However, if one were to reproduce, for example, a Hamiltonian containing a hopping term like:
\begin{eqnarray}
\mathcal H = \sum_i \omega \, \hat{\alpha}_i^\dagger \hat{\alpha}_i - J \sum_{i > j} (e^{i\varphi}  \hat{\alpha}_{i}^\dagger \hat{\alpha}_j + \text{ H.c.}),\label{eq:reciprocal_coupling}
\end{eqnarray}
then a feedback
involving inter-oscillator forces
is needed. In the equation above, $J$ is the hopping amplitude, and $\varphi$ is the tunnelling phase that mimics the Peierls phase gained by a charged particle hopping on a tight-binding lattice in the presence of a magnetic vector potential. Hence, the latter may be designed to engineer artificial magnetic fields and  topological quantum Hall tight-binding models in larger systems. 

In order to realise Eq.~\ref{eq:reciprocal_coupling}, we require:
\begin{eqnarray*}
{F}_{i j}^{\text{RWA}} \alpha_j  = \begin{cases}
0 & \text{for } i=j\\
\sqrt{2 \omega}J \,e^{i \varphi_{j,i}} \alpha_j
 & \text{for } i\neq j
\end{cases}
\end{eqnarray*}
with
$\varphi_{j,i} = \varphi$ and $\varphi_{i,j} = -\varphi$. In this case, a possible choice of feedback applied to the $i$-th oscillator depending on the $j$-th one is:
\begin{eqnarray*}
F_i &=& A_{j, i} X_j + B_{j, i} P_j,\quad i\neq j \\
&=& A_{j, i} \frac 1{\sqrt{2\omega}} (\alpha_j + \alpha_j^*) - i\, B_{j, i} \sqrt{\frac \omega 2}\, (\alpha_j-\alpha_j^*).
\end{eqnarray*}
Hence, one needs to set $A_{j,i} = 2 J\, \omega \cos\varphi_{j,i}$ and $B_{j,i} = - \,2 J \sin \varphi_{j,i}$. Experimentally, this can be realised by measuring both jerk and acceleration from one oscillator, and applying a commensurate feedback to the other. Note that for a system with many oscillators, the hopping amplitudes and phases can be chosen to have arbitrary spatial dependence so as to encode the desired lattice geometry, connectivity and gauge fields.

In experiment, working with normalized measurements for $\tilde{x}$ and $\tilde{p}$, the application of conjugate forces for the implementation of complex hopping is relatively straightforward. We demonstrate this control in Fig.~\ref{FIG:fig3}. Figure~\pref{FIG:fig3}{a} plots the dynamics of the mechanical energy stored in oscillator 1 (red) and oscillator 2 (blue) as a function of evolution time under the applied inter-oscillator feedback. Prior to the plotted dynamics, we again include an initialization step during which energy is deposited into oscillator 1 by a sinusoidal drive. High-visibility Rabi oscillations are observed in the mechanical energy dynamics, with energy flowing from oscillator 1 to oscillator 2, and back. As noted by the inset, these dynamics occur for a tunneling phase of $\varphi = \pi$.

To see the consequence of the imposed tunneling phase $\varphi$, we need only look at the traces of either $\tilde{x}$ or $\tilde{p}$ for the two oscillators, which provide information about the \textit{relative phase} of the oscillators. Figure~\pref{FIG:fig3}{b} shows the dynamics of the position signals of the two oscillators, with a zoom in on the time (near 10~s) at which the two oscillators first have nearly equal mechanical energies. For the case of $\varphi = \pi$, we have a position-dependent force, and the two signals are out of phase by $\pi/2$, similar to the usual physical scenario of coupled oscillating masses. However, unlike the usual case, we find that as energy moves from oscillator 1 to 2, the $\tilde{x}_2$ signal actually leads the $\tilde{x}_1$ signal, whereas it would normally lag for a physical spring. This reflects our implementation of a $\pi$ tunneling phase. The lower two panels indicate the consequence of introducing a momentum-dependence to the inter-oscillator coupling. In particular, for $\varphi = -\pi/2$, the coupling terms become fully momentum-dependent, leading to dynamics in which the oscillators swing in phase as energy is transferred, signaling a concomitant
change to the phase structure of the system's eigenstates.

\begin{figure*}[t]
	\includegraphics[width=1.95\columnwidth]{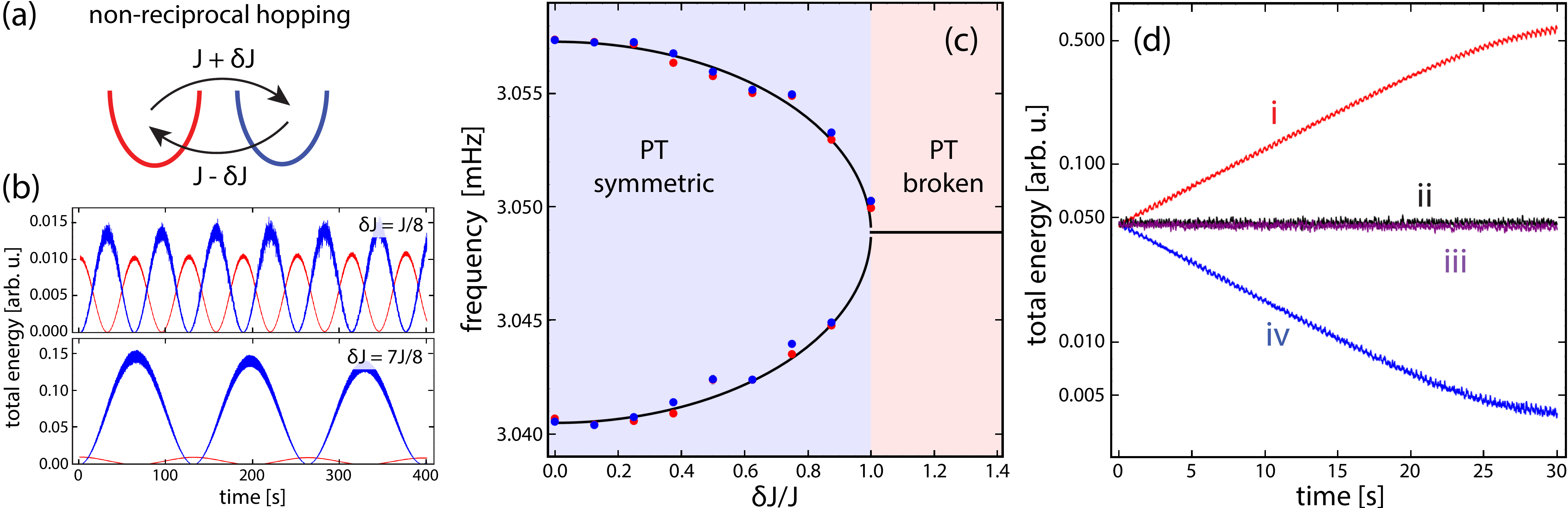}
	\centering
	\caption{\label{FIG:fig5}
		\textbf{Non-reciprocal hopping and PT symmetry breaking}
        \textbf{(a)}~Two oscillators coupled with non-reciprocal hopping terms, $J \pm \delta J$.
        \textbf{(b)}~Experimental population dynamics, starting with all energy in the first oscillator, for $\delta J / J = 0.125$ and $\delta J / J = 0.875$.
        \textbf{(c)}~Fit-extracted peak locations of the power spectra based on experimental dynamics when starting with all energy in oscillator 1 (red dots) and with all energy in oscillator 2 (blue dots). The solid lines relate to the expected real eigenspectrum for $f_0 = 3.0489$~Hz and $J = 2\pi \times 8.4$~mHz.
        Error bars, which are smaller than the points, relate to the standard error of the fits to two Gaussian peaks.
        \textbf{(d)}~Experimental total energy ($E_1 + E_2$) dynamics for excitation of eigenmodes in the case of fully asymmetric hopping (i and iv, with $J = 0$) and fully symmetric hopping (ii and iii, with $\delta J = 0$). For the symmetric case, the initial states relate to the two oscillators having equal energy and an initial relative oscillator phase of $0$ (ii) and $\pi$ (iii). For the asymmetric case, the initial states relate to the two oscillators having equal energy and an initial relative oscillator phase of $\pi/2$ (i) and $-\pi/2$ (iv).
	}
\end{figure*}

To note, there are slight differences in the relative amplitudes of the red and blue curves during this time window for the three different panels of Fig.~\pref{FIG:fig3}{b}, relating to slightly different hopping rates for the different $\varphi$ values. This stems from the presence of a small \textit{natural} coupling between the oscillators, which do in fact share a common physical support apparatus for convenience. To avoid such effects, one can simply actively cancel such natural contributions prior to adding synthesized hopping (which is done later for the probing of non-reciprocal hopping in Fig.~\ref{FIG:fig5}). Here, by working with synthesized coupling terms that are much larger than the natural coupling, we simply work above it and tolerate a small $\varphi$-dependence to the hopping rate.

Figure~\pref{FIG:fig3}{c} depicts more comprehensively how the relative phase and mechanical energy of the oscillators evolve throughout the dynamics. Taking advantage of the ability to simultaneously measure both the $\tilde{x}$ and $\tilde{p}$ signals at all sites, we reconstruct both the normalized energy imbalance of the two oscillators, $z = (E_1 - E_2) / (E_1 + E_2)$, as well as the relative phase between them, $\Delta \phi = \phi_1 - \phi_2$, where we simply determine the energies and phases of the oscillators as $E = (\tilde{x}^2 + \tilde{p}^2)/2$ and $\phi = \textrm{arg}(\tilde{x} + i \tilde{p})$. The evolution of $z$ and $\Delta \phi$ under the system evolution maps out the phase-space dynamics of this simple two-mode system. For the case of hopping with equal site energies, these dynamics should map out trajectories as shown in Fig.~\pref{FIG:fig3}{c}, with full oscillations between $z = \pm 1$, and with the curves centered about a $\Delta \phi$ value determined directly by the tunneling phase $\varphi$. In an analogous description in terms of an effective Bloch sphere (with the energy initiated at site 1 relating to a state vector initially aligned along $+z$), these trajectories simply relate to the paths traversed in a response to different applied tunneling ``torque'' vectors lying (at azimuthal angles $\varphi$) in the equatorial plane.

We now integrate our demonstrated control over site energy terms with this control over inter-site hopping. In particular, we investigate how the frequency spectrum of this two-oscillator system evolves as we add a variable inter-site frequency offset $\Delta$ (applied as equal amplitude shifts of $\pm \Delta/2$ to oscillators 1 and 2, respectively) while keeping a fixed inter-oscillator hopping rate $J$, as depicted in Fig.~\pref{FIG:fig4}{a}. Similar to Fig.~\ref{FIG:fig2}, we obtain these frequency spectra by looking at the dynamics and performing a numerical Fourier transform. Figure~\pref{FIG:fig4}{b} shows this spectrum, with the canonical avoided crossing encountered near $\Delta = 0$ as the modes of the two oscillators hybridize due to the engineered hopping. To note, the color indexing of Fig.~\pref{FIG:fig4}{b} reflects the weight of the response at a given frequency that is found to relate to energy in oscillator 1 (red) and oscillator 2 (blue). Specifically, we investigate the dynamical evolution following the initialization of energy at site 1 or 2, yielding distinct Fourier spectra as depicted in red and blue, respectively. One sees that for large $\Delta$ the eigenmodes are nearly localized to the individual oscillators, while they are near-evenly delocalized near resonance. Finally, in Fig.~\pref{FIG:fig4}{c}, starting with energy in oscillator 1, we plot the dynamical evolution of the normalized energy in oscillator 2 as a function of $\Delta$. This Chevron Rabi pattern relates to high-visibility oscillations near $\Delta = 0$, with faster and lower-amplitude oscillations for larger values of the site-to-site frequency mismatch.


\subsection{Non-reciprocal Inter-oscillator Hopping}

Going further, one advantage of using feedback to engineer inter-oscillator hopping terms is that it is straightforward to realise non-reciprocal couplings.
Non-reciprocal Hamiltonians~\cite{Vitelli-NonRecip} naturally host chiral phenomena, and are intimately connected to the physics of non-Hermitian mechanics~\cite{Ueda-NH-review}.
While non-reciprocity would be challenging to access in physical systems governed by Newtonian mechanics -- where forces come in equal and opposite pairs -- it can effectively be engineered in active matter~\cite{shankar2020topological,Brandenbourger2019,Coulais-PNAS} and in systems featuring dissipation~\cite{Bo-NonRecip}.
Here, as we show, it can be engineered at will in synthetically coupled mechanical networks.

We consider the following Hamiltonian:
\begin{eqnarray*}
\mathcal H = \sum_i \omega \, \hat{\alpha}_i^\dagger \hat{\alpha}_i - \sum_{i>j} [(J+\delta J) e^{i\varphi} \hat{\alpha}_i^\dagger \hat{\alpha}_j +\\
(J-\delta J) e^{-i\varphi} \hat{\alpha}_j^\dagger \hat{\alpha}_i], 
\end{eqnarray*}
where the change in the hopping amplitude $ J \rightarrow J \pm\delta J$  [\textit{cf.}~$\mathcal H$ in Eq.~\ref{eq:reciprocal_coupling}] represents a non-reciprocal inter-oscillator coupling.  This type of non-reciprocal coupling is well-known from the Hatano-Nelson model~\cite{HN-PRL-96,HN-PRB-97,HN-PRB-98}, and it not only selects a preferential direction of hopping, but more importantly causes this Hamiltonian to become non-Hermitian ($\mathcal H^\dagger \neq \mathcal H$). Note that for 
$|\delta J / J| < 1$, this Hamiltonian is $\mathcal{PT}$-symmetric and its eigenvalues are real. Conversely, if
$|\delta J / J| > 1$,
the $\mathcal{PT}$-symmetry of the system is broken and the energies become complex. At $|\delta J / J| = 1$ the energy levels coalesce in an exceptional point and the $\mathcal{PT}$ phase transition takes place.
Explicitly, the non-reciprocal double-well (with coupled left and right modes, $|L\rangle$ and $|R\rangle$) possesses eigenmodes $\pm \sqrt{J^2 - \delta J^2} / (\delta J - J)|L\rangle + |R\rangle$ (up to normalization factors), having eigenfrequencies $\pm \sqrt{J^2 - \delta J^2}$, respectively.

Following the earlier discussed procedure, it is straightforward to obtain the new definitions for $A_{j,i}$ and $B_{j,i}$:
\begin{eqnarray*}
A_{1,2} &=& 2 (J+ \delta J)\, \omega \cos\varphi \; \text{ and }\; B_{1,2} = - \,2 (J+ \delta J) \sin \varphi, \\
A_{2,1} &=& 2 (J- \delta J)\, \omega \cos\varphi \; \text{ and }\; B_{2,1} = \,2 (J- \delta J) \sin \varphi;
\end{eqnarray*}
with analogous physical meaning as before. Because each of these force terms are added by hand, non-reciprocal hoppings are as natural to engineer in this setting as reciprocal ones.

In Fig.~\ref{FIG:fig5} we experimentally demonstrate the ability to engineer non-reciprocal hopping terms (depicted in the cartoon of Fig.~\pref{FIG:fig5}{a}), driving this two-site system across a  $\mathcal{PT}$ symmetry-breaking phase transition. To note, while in the earlier data (Figs.~\ref{FIG:fig2}-\ref{FIG:fig4}) there existed an additional, small contribution to the hopping due to physical coupling (via the shared support structure), for this data we take care to first actively cancel, via feedback, the physical coupling (scale $\lesssim 1$~mHz) prior to introducing our synthetic coupling forces.

For a fixed value of $J \approx 2\pi \times 8.4$~mHz, we introduce a tunable hopping asymmetry $\delta J$. Figure~\pref{FIG:fig5}{b} shows the dynamics of the measured oscillator energies, beginning with energy residing in oscillator 1, for values of $\delta J = J/8$ (top) and $7J/8$ (bottom). Unlike in the case of reciprocal hopping terms, the total mechanical energy is not fixed throughout these dynamics, even in the $\mathcal{PT}$-symmetric phase. As energy flows from oscillator 1 to 2, the energy of the system grows. The difference of scale of the two panels of Fig.~\pref{FIG:fig5}{b} reflects the fact that this growth in energy becomes more pronounced for larger values of $\delta J$ approaching the $\mathcal{PT}$ symmetry-breaking phase transition.
As stated above, the eigenstates of the system in the $\mathcal{PT}$-symmetric regime similarly reflect an asymmetry between left and right modes, which for larger lattices gives rise to the non-Hermitian skin effect~\cite{Non-Herm-Skin}.
One additionally finds a clear difference in the rate of population exchange (Rabi) dynamics observed in these two cases, with the dynamics slowing down appreciably as $\delta J/J$ approaches 1. This slow-down of the dynamics reflects a change to the eigenspectrum of the system, with the real eigenvalues coalescing at an exceptional point at $\delta J = J$, then transforming into imaginary eigenenergies, reflecting modes that undergo pure exponential decay or gain (ignoring the intra-envelope $\tilde{x}$ or $\tilde{p}$ dynamics at the bare frequency $\omega$).

We can again directly investigate this frequency response of the system upon approaching the $\mathcal{PT}$-breaking phase transition by simply Fourier-transforming the oscillation dynamics (taken over 400~s) and extracting a peak or peaks in the resulting power spectrum. We perform this analysis for both the cases of starting in oscillator 1 (red points) and oscillator 2 (blue points) and plot the resulting resonance values in Fig.~\pref{FIG:fig5}{c}. We perform this analysis solely in the $\mathcal{PT}$ symmetric regime (and up to $\delta J = J)$, but find a clear closing of the energy gap in the system as the $\mathcal{PT}$ phase transition is approached. The measured frequency resonance values are in good agreement with the plotted theory curve, which relates to the form $f_0 \pm (J/2\pi)\sqrt{ 1 - (\delta J /J)^2}$ with $f_0 = 3.0489$~Hz and $J = 2\pi \times 8.4$~mHz.

To gain insight into the structure of the eigenmodes in the $\mathcal{PT}$-broken region, we seek to prepare these eigenmodes directly and observe their evolution. Figure~\pref{FIG:fig5}{d} contrasts the behavior of prepared eigenmodes in the purely reciprocal case (curves ii and iii) to those in the purely non-reciprocal case (i and iv). In the reciprocal case ($\delta J = 0$, $J \approx 2\pi \times 8.4$~mHz), we prepare both in-phase and out-of-phase eigenmodes of the coupled two-oscillator system, by initially driving the two oscillators (to deposit mechanical energy) with a controlled phase difference, to result in equal-energy superposition states with relative phases of $0$ (ii) and $\pi$ (iii). We plot, in Fig.~\pref{FIG:fig5}{d}, the dynamics of the total energy $E = E_1 + E_2$ for these two cases, observing no appreciable variation in this reciprocal hopping scenario.

For the purely non-reciprocal case ($J = 0$, $\delta J \approx 2\pi \times 8.4$~mHz), the eigenstate structure is maximally distinct to the reciprocal case, yielding equal-energy superposition states with relative phases of $\pm \pi/2$. Using the same general initialization procedure, we prepare these eigenmodes of the non-reciprocal double-well, and we indeed observe distinct dynamics of the total energy, resulting in an exponential growth of the mode with relative phase $\pi/2$ (i) and an exponential attenuation of the mode with relative phase $-\pi/2$ (iv), at least up until the scale at which uncontrolled nonlinearities modify this picture.

\begin{figure*}[t!]
	\includegraphics[width=1.99\columnwidth]{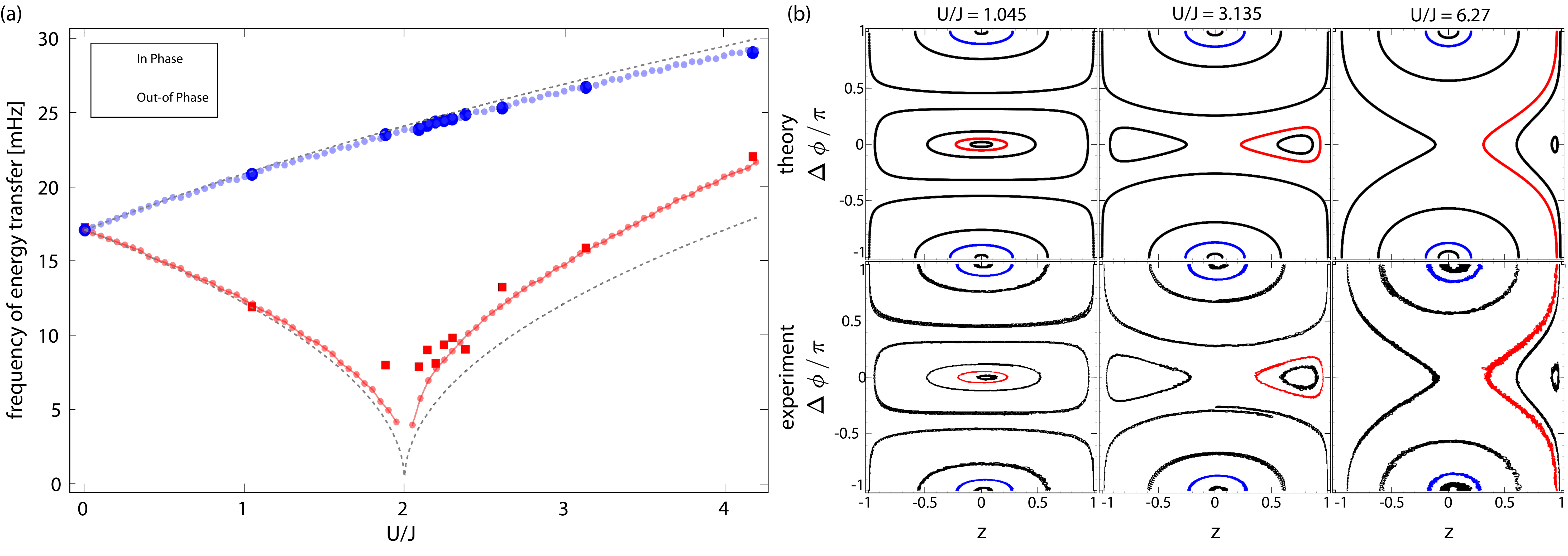}
	\centering
	\caption{\label{FIG:fig6}
		\textbf{Eigenmodes and phase-space dynamics of a two-oscillator system with synthetic interactions.}
        \textbf{(a)}~Mode frequencies for a double-well system with synthetic Hartree-like local interactions, implemented via self-feedback forces $F_i \propto E_i \tilde{x}_i$, where $E_i$ is the total mechanical energy in mode $i$.
        The frequencies are determined by fits to the experimental population imbalance dynamics of in-phase ($\Delta\phi = 0$) and out-of-phase ($\Delta\phi = \pi$) modes with an initial population imbalance $z=0.3$. The out-of-phase mode stiffens with increasing $U/J$, while the in-phase mode softens across the self-trapping transition at $U/J \sim 2$. The dark blue circles (dark red squares) relate to experimental data, the light blue (red) points relate to numerical simulations, and the light gray dashed lines relate to the approximate analytical expressions described in the text.
        \textbf{(b)}~Phase-space maps of the experimental dynamics of the relative oscillator phase $\Delta \phi$ and normalized energy imbalance $z$, for $U/J \approx 1$, 3, and 6. To note, small residual loss and/or gain terms leads to slightly disconnected trajectories in the experimental panels.
	}
\end{figure*}

\subsection{Synthetic Interactions}

Finally, it is worth considering the outcome of weak non-linear feedback, as this is a way to simulate mean-field interactions in the Hamiltonian mapping.
In the context of classical emulation experiments, this feedback based approach has connections to recent proposals and realizations in photonic~\cite{Duncan2020} and electronic~\cite{Khanikaev-nonlinear} networks.
As an example of how feedback can introduce synthetic interactions, consider an energy-dependent feedback $F_i = (A_{i,i} X_i^2 + B_{i,i}  P_i^2) X_i$. In this case, a measure related to the ``energy'' of the system is used to create a non-linear on-site potential. By substituting the results in Eq.~\ref{eq:XP} into the form of the feedback, we have
\begin{eqnarray*}
F_i &= &
\bigg[\tilde{A}_{i,i} ({\alpha_i} + {\alpha_i^*})^2 - \tilde{B}_{i,i} ({\alpha_i} - {\alpha_i^*})^2\bigg] (\alpha_i + \alpha_i^*)\\
&=& \Omega_{-,i} ({\alpha_i}^3 + {\alpha_i^*}^3 + {\alpha_i}^2 \alpha_i^* + {\alpha_i^*}^2 \alpha_i) \\
&&+ \Omega_{+,i} ({\alpha_i}^2 \alpha_i^* + {\alpha_i^*}^2 \alpha_i),
\end{eqnarray*}
where $2 \omega \tilde{A}_{i,i} = {A}_{i,i}$, $2\tilde{B}_{i,i} = {B}_{i,i}$ and $\Omega_{\pm, i} = \tilde{A}_{i,i}\pm \tilde{B}_{i,i}$. When applying the RWA one should be careful in neglecting terms, since also the contribution from terms like ${\alpha_i}^3$ and  $ {\alpha_i^*}^2 \alpha_i$ can be discarded. In the former case, we have a much faster co-rotating wave, while in the latter case the net contribution in terms of frequency is $\approx e^{-i\omega t}$, which is counter-rotating with frequency $\approx 2 \omega$ in the rotating frame. We are then left with:
\begin{eqnarray*}
{F}_{i j}^{\text{RWA}} \alpha_j  = \begin{cases}
2 \tilde{A}_{i,i} \,|{\alpha_i}|^2 \alpha_i & \text{for } i=j\\
0
 & \text{for } i\neq j
\end{cases}
\end{eqnarray*}
Then the Hamiltonian in Eq.~\ref{eq:H} follows as:
\begin{eqnarray*}
\mathcal H =\sum_i \omega \hat{\alpha}_i^\dagger \hat{\alpha}_i - \sum_{i} U_i \hat{\alpha}_i^\dagger \hat{\alpha}_i^\dagger \hat{\alpha}_i \hat{\alpha}_i, \quad U_i = \tilde{A}_{i,i}/\sqrt{2 \omega} , 
\end{eqnarray*}
where the second term is the on-site interaction.
When written in terms of the equation of motion, Eq.~\ref{eq:RWA}, after replacing operators by c-numbers, features a term that reflects a local (diagonal, Hartree) mean-field interaction.

We now implement this local mean-field interaction in our two-oscillator double-well, with a common value of the nonlinear $U$ term applied to each oscillator.
Here, the term $U$ describes a local angular frequency shift due to mean-field nonlinearity, where the scale is such that it represents the shift that would be experienced by a given oscillator when all of the mechanical energy resides in said oscillator (with total mechanical energy approximately conserved in the following scenarios we explore).
We calibrate the $U$ term for each oscillator by directly measuring the oscillator frequency in an uncoupled configuration for several values of the nonlinear feedback coefficient as well as several values of the mechanical energy. To note, while the engineered variations of the quartic nonlinearity are indeed found to be linearly tunable by application of feedback force terms $\propto (\tilde{x}_i^2 + \tilde{p}_i^2) \tilde{x}_i$, our calibrations additionally account for \textit{natural} contributions to the oscillator nonlinearity due to the physical properties of the springs. For example, to achieve a ``non-interacting'' scenario, we in fact first cancel the relatively weak quartic nonlinearities that occur naturally due to the physical properties of our springs. Controlled nonlinear terms are then added to this ``non-interacting'' starting point. To note, this cancellation was performed for the study presented earlier in Fig.~\ref{FIG:fig5}, and for all studies presented hereafter.

We now use our ability to apply a tunable nonlinearity to a tunnel-coupled system to explore the self-trapping phase transition of the canonical Josephson double-well~\cite{Raghavan,Maciej-self-trap}.
This mean-field model describing interacting bosonic excitations in a coherently coupled two-mode system is ubiquitous, describing physical systems ranging from
polariton fluids~\cite{Bloch-polariton} to cold atomic gases~\cite{Albiez-DirectJosephson,Levy2007,Anker-SelfTrap,Zibold} to classical photonics~\cite{chiao-selftrap,tappert,gordon,Snyder:91}.
Working with a fixed hopping $J \approx - 2\pi\times 8.4$~mHz, we plot in Fig.~\pref{FIG:fig6}{a} the mode spectrum of this double-well system as we tune the nonlinear $U$ term. Specifically, we present the measured frequency dependence of prepared in-phase ($\Delta \phi = 0$) and out-of-phase ($\Delta \phi = \pi$) superposition modes, each having a small initial inter-well energy imbalance of $z = 0.3$. Similar to before, the relative phase $\Delta \phi$ and energy imbalance $z$ of the initial state is controlled by the relative amplitudes and phases of the applied sinusoidal drives during an initial preparation time step. Here, we determine the mode frequencies of the system by simply taking the inverse of the time separation between local extrema of the $z$ dynamics.

\begin{figure*}[t!]
	\includegraphics[width=1.85\columnwidth]{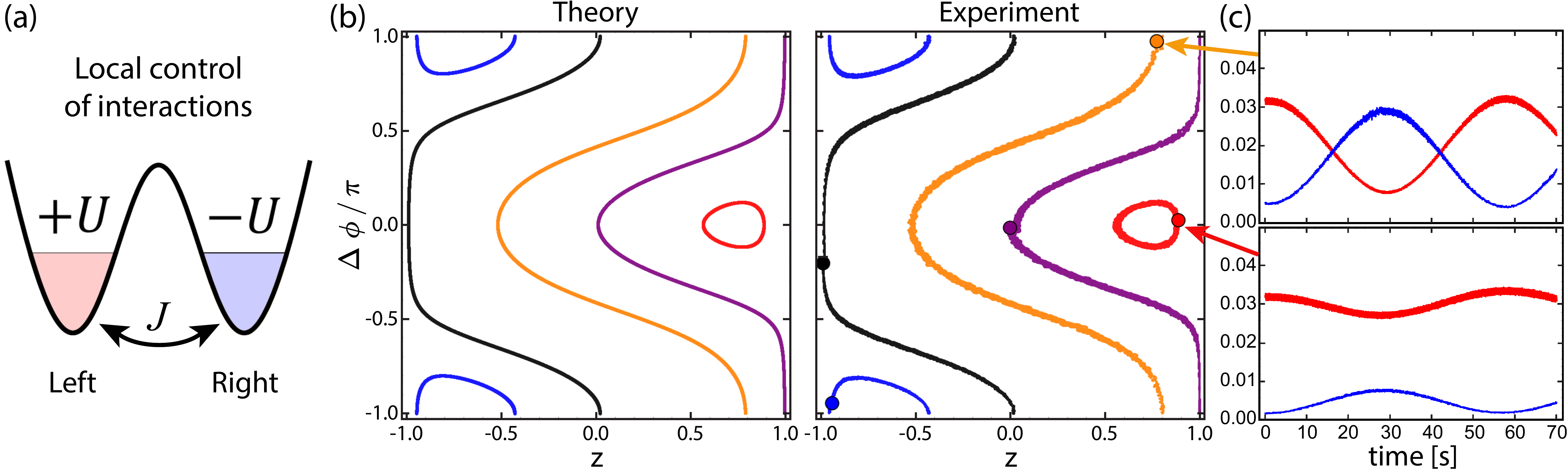}
	\centering
	\caption{\label{FIG:fig7}
		\textbf{Phase-space dynamics in a two-oscillator system with oscillator-dependent interactions.}
        \textbf{(a)}~Cartoon depiction of two-oscillator system mimicking a tunnel-coupled double well with well-dependent nonlinear interactions.
        \textbf{(b)}~Left: theoretical phase-space portrait of relative phase and population in the double-well for $U = 2.2 J$ , revealing self-trapped trajectories (blue and red) and modes with full population swings (orange).  Right: experimental phase-space dynamics starting from different initial conditions (indicated by colored circles). \textbf{(c)}~Dynamics of the measured
        mechanical energy in the left and right oscillators (shown as red and blue signals, respectively, with arbitrary units), for the indicated experimental trajectories.
	}
\end{figure*}

Specifically, in Fig.~\pref{FIG:fig6}{a} we show how the measured in-phase and out-of-phase mode frequencies evolve as we tune the ratio of $U/J$ across the expected self-trapping phase transition at $U = 2J$. For zero quartic nonlinearity, we find excellent agreement between the measured frequencies of the in-phase and out-of-phase modes. Upon adding a weak $U$ term, however, these modes undergo radically different responses. The out-of-phase or $\pi$ mode becomes stiffened, seeing its frequency increase directly as $U$ is increased. The in-phase mode, in contrast, acquires a decreased mode frequency, leading to very slow dynamics near $U = 2J$. Beyond $U = 2J$, the frequency of the in-phase mode begins to increase, approaching that of the stiffened $\Delta \phi = \pi$ mode.

This observed difference in the modal frequency response -- that the $\pi$ mode is continuously stiffened for increasing $U$, while the $0$-phase mode undergoes mode-softening -- is completely in line with the expected response associated with the Josephson double-well~\footnote{However, the roles of the two modes are reversed, as we operate with a negative $J$ value (effectively a $\pi$ tunneling phase). In the canonical bosonic Josephson junction~\cite{Raghavan}, the out-of-phase $\pi$-mode undergoes mode-softening while the in-phase plasma mode is purely stiffened.} and its supported dynamical self-trapping phase transition~\cite{Raghavan,Maciej-self-trap}. Our measured mode frequencies are in fair agreement with approximate analytical forms for the 0-phase (plasma) mode and $\pi$ mode oscillation frequencies (Eqs.~4.7 and 4.10 of Ref.~\cite{Raghavan}, respectively), which we plot as dashed black lines. We note some disagreement between our data and these analytical formulae, however, which is expected as these expressions are strictly valid only near $z = 0$.
We find considerably better agreement between our data and the frequencies predicted by numerical simulations of an ideal Josephson double well for our given initial $z$ value of 0.3, plotted as the connected-dot curves of Fig.~\pref{FIG:fig6}{a}.

There exist further distinctions between the response of these two modes beyond their frequency behavior. Specifically, while the frequencies of the in-phase mode are similarly small on either side of the mode-softening encountered at $U = 2J$, we find that there are net swings of the energy imbalance $z$ for values $U < 2J$, \textit{i.e.}, $z$ changes sign during the dynamics. In contrast, for $U > 2J$ we find that the excess mechanical energy becomes \textit{self-trapped} to the oscillator in which it is initiated. This self-trapping transition of the in-phase mode reflects a drastic modification of the phase space portraits describing this system, as a separatrix moves through the phase space for increasing $U/J$~\cite{Raghavan,Maciej-self-trap}.

We now examine this behavior in Fig.~\pref{FIG:fig6}{b} by mapping out the phase space portraits for several values of $U/J$, utilizing the same approach as in Fig.~\pref{FIG:fig3}{c}.
The out-of-phase modes, indicated by the blue trajectories, remain relatively unaffected in the three phase-space portraits for $U/J$ values of $\sim$1, 3, and 6. In contrast, trajectories starting near the unstable fixed point at $z = 0$ and $\Delta \phi = 0$ are significantly altered by the introduction of nonlinear $U$ term. These trajectories first become self-trapped in both relative phase $\Delta \phi$ and energy imbalance $z$ for moderate $U$ values just beyond the self-trapping transition (\textit{i.e.}, for $U \sim 3J$). For still larger values, we encounter a different form of self-trapped mode, having $z$ values confined to either side of $z = 0$ but experiencing a running relative phase $\Delta \phi$.
These results, in excellent agreement with theory, demonstrate how the incorporation of real-time measurements can enable both the implementation of synthetic interactions as well as visualization of the resulting nonlinear dynamics.

While the mean-field Josephson model with uniform interactions is naturally realized by a range of physical systems, our general approach to engineering synthetic nonlinearities allows for the extension to more 
intricate forms of interactions. As one simple demonstration of this, we examine in Fig.~\ref{FIG:fig7} the phase-space dynamics of a double-well system under the application of a well-dependent interaction term $U_1 = -U_2 = 2.2 J$. We again start from different points in phase space (indicated by the colored disks) and study the phase-space dynamics of $z$ and $\Delta \phi$. Compared to the double-well with uniform quartic nonlinearity, the phase space map is significantly altered, showing two distinct self-trapped regions at relative phases of $\Delta \phi = 0$ and $\pi$. We again find excellent agreement between the observed trajectories and those predicted by theory.

%
%
%

\subsection{Additional Possible Terms}

Having given specific examples of the terms that we experimentally engineer in this work, we now briefly highlight, from a theoretical view-point, some general considerations for the type of terms that could be accessible with this approach [\textit{cf}.~Eq.~\ref{eq:H}]. 
Firstly, an advantage of using feedback to engineer the system is that the resulting Hamiltonian can be arbitrarily long-ranged or spatially-structured when it is scaled up to include more oscillators. For example, there are no fundamental theoretical constraints on the range or type of interactions and couplings that are possible between different oscillators, which allows for, in principle, arbitrary network connectivity and nonlinearities, provided that the feedback is weak. 

Secondly, we have assumed above that the form of the feedback is time-independent, such that the resulting Hamiltonian is also time-independent. However, this is not required for the RWA to hold so long as the natural frequency, $\omega$, remains the largest frequency scale in the problem.
This allows, \textit{e.g.}, for the investigation of time-modulated Hamiltonians, to introduce still further control for engineering effective models~\cite{Salerno-alpha,Salerno_2014_epl} or for the investigation of Floquet or stroboscopic Hamiltonians for their own sake~\cite{floquet-bukov,casati}.
Alternatively, slow variations of the feedback parameters can enable the exploration of adiabatic geometric response and dynamical phase transitions, as well as the population and exploration of system eigenstates by Hamiltonian annealing.

Thirdly, as discussed above, the RWA reduces the feedback to $F_i \!\rightarrow\! \sum_{j}{F}_{ij}^{\text{RWA}} \alpha_j $ by only keeping co-rotating terms with a frequency $\approx \omega$ in the lab-frame. In general, such terms must have the number of $\alpha$ factors being one greater than the number of $\alpha^*$ factors, as any other combinations should average to zero in the $\omega \rightarrow \infty$ limit. This in turn means that in the RWA Hamiltonian [Eq.~\ref{eq:H}] the only terms present will contain equal numbers of $\alpha$ and $\alpha^*$ variables. In the language of quantum creation and annihilation operators, this corresponds to Hamiltonians which conserve the total number of particles in the system, but do not necessarily constrain the local particle number or the energy of the system. This therefore allows for many other types of terms, such as correlated hopping terms in which, for example, two particles hop at once.
Considering only two-body (quartic) interactions, the degree of control afforded by this approach should have direct applications in the engineering of interacting Hamiltonians with exotic forms~\cite{Anderson-molecule} or fine-tuned symmetries~\cite{pretko-circuit,Jendrz-2019}.
This approach even allows for the direct engineering of Hamiltonians dominated by beyond-quartic terms~\cite{mintert-3body}, which could enable the realization of exotic phases requiring higher-order interactions, without complications due to the presence of lower-order terms~\cite{lapa-int-enab}.

\subsection{Beyond the RWA}

Finally, it is worth noting that the RWA is strictly valid for weak feedback in the limit that $\omega \rightarrow \infty$. In general, there will be corrections to the RWA coming from the coupling between the $\alpha$ and $\alpha^*$ variables in the dynamical equations [\textit{cf.}~Eq.~\ref{eq:dynamics}]. For example, as discussed in  Refs.~\cite{Salerno-alpha,Salerno_2014_epl}, this can lead to corrections such as, at lowest order, an overall shift to the resonance frequency, which is a classical analogue of the Bloch-Siegert shift. 

At even lower frequencies or for stronger feedback, the RWA cannot be applied and counter-rotating feedback terms, such as those linearly-dependent only on $\alpha_i^*$, should not be neglected. In such cases, by a similar line of reasoning to that above, the system can be mapped to a tight-binding Hamiltonian that now does not conserve the total particle number. As is well-known, such particle-non-conserving terms in bosonic Hamiltonians can lead to parametric instabilities, as recently studied, for example, in topological models~\cite{Ohe-ChiralMagnonic,Peano2016,Brandes-Bog,Salerno-alpha,Salerno_2014_epl}.


$\ $

\section{Conclusion}
\label{conc}

We have presented a general recipe for engineering synthetic lattice tight-binding models, featuring both linear and nonlinear terms, based on the real-time measurement of and feedback on arrays of isolated mechanical oscillators. We have experimentally demonstrated the basic elements of this approach, including the control of real and imaginary site energies, complex and non-reciprocal inter-site hopping terms, and engineered nonlinear interactions. The presented approach is general, directly applicable to classical emulators based on mechanics or electronics and photonics~\cite{top-phot}, but also intimately related to recent ideas for the steering of quantum systems into new many-body phases via measurement-based feedback~\cite{SpielmanFeedback1,SpielmanFeedback2,DeutschFeedback1,DeutschFeedback2}.
Looking forward, as this approach is expanded to many-site arrays of classical oscillators, the ability to synthesize near-arbitrary mean-field lattice Hamiltonians will open up new opportunities for exploring exotic transport phenomena in a highly accessible laboratory setting. 
More generally, the incorporation of measurement-aided approaches
promises to expand our capabilities for synthetic lattice and synthetic dimensions~\cite{Review-OzawaPrice} engineering.

\section{Acknowledgements}
We thank Camelia Prodan for stimulating discussions.
This material is based upon work supported by the National Science Foundation under grant No.~1945031. C.~B.-M. and E.~C. acknowledge REU support from the National Science Foundation under grant No.~1950744.
R.~A., Y.~S., and M.~C. acknowledge support by the Philip J. and Betty M. Anthony Undergraduate Research Award of the UIUC Department of Physics. R.~A. acknowledges support by the Lorella M. Jones Undergraduate Research Award of the UIUC Department of Physics. Y.~S. acknowledges support by the Jeremiah D. Sullivan Undergraduate Research Award of the UIUC Department of Physics.
T.~O. acknowledges support from JSPS KAKENHI Grant No. JP20H01845, JST PRESTO Grant No. JPMJPR19L2, JST CREST Go. Number JPMJCR19T1, and RIKEN iTHEMS. E.~M. and H.~M.~P. are supported by the Royal Society via grants UF160112, RGF\textbackslash EA\textbackslash 180121 and RGF\textbackslash R1\textbackslash 180071.

\bibliographystyle{apsrev4-1}
\bibliography{SynthOscBib}

\begin{thebibliography}{65}%
\makeatletter
\providecommand \@ifxundefined [1]{%
 \@ifx{#1\undefined}
}%
\providecommand \@ifnum [1]{%
 \ifnum #1\expandafter \@firstoftwo
 \else \expandafter \@secondoftwo
 \fi
}%
\providecommand \@ifx [1]{%
 \ifx #1\expandafter \@firstoftwo
 \else \expandafter \@secondoftwo
 \fi
}%
\providecommand \natexlab [1]{#1}%
\providecommand \enquote  [1]{``#1''}%
\providecommand \bibnamefont  [1]{#1}%
\providecommand \bibfnamefont [1]{#1}%
\providecommand \citenamefont [1]{#1}%
\providecommand \href@noop [0]{\@secondoftwo}%
\providecommand \href [0]{\begingroup \@sanitize@url \@href}%
\providecommand \@href[1]{\@@startlink{#1}\@@href}%
\providecommand \@@href[1]{\endgroup#1\@@endlink}%
\providecommand \@sanitize@url [0]{\catcode `\\12\catcode `\$12\catcode
  `\&12\catcode `\#12\catcode `\^12\catcode `\_12\catcode `\%12\relax}%
\providecommand \@@startlink[1]{}%
\providecommand \@@endlink[0]{}%
\providecommand \url  [0]{\begingroup\@sanitize@url \@url }%
\providecommand \@url [1]{\endgroup\@href {#1}{\urlprefix }}%
\providecommand \urlprefix  [0]{URL }%
\providecommand \Eprint [0]{\href }%
\providecommand \doibase [0]{http://dx.doi.org/}%
\providecommand \selectlanguage [0]{\@gobble}%
\providecommand \bibinfo  [0]{\@secondoftwo}%
\providecommand \bibfield  [0]{\@secondoftwo}%
\providecommand \translation [1]{[#1]}%
\providecommand \BibitemOpen [0]{}%
\providecommand \bibitemStop [0]{}%
\providecommand \bibitemNoStop [0]{.\EOS\space}%
\providecommand \EOS [0]{\spacefactor3000\relax}%
\providecommand \BibitemShut  [1]{\csname bibitem#1\endcsname}%
\let\auto@bib@innerbib\@empty
\bibitem [{\citenamefont {Debye}(1912)}]{Debye}%
  \BibitemOpen
  \bibfield  {author} {\bibinfo {author} {\bibfnamefont {P.}~\bibnamefont
  {Debye}},\ }\href {\doibase https://doi.org/10.1002/andp.19123441404}
  {\bibfield  {journal} {\bibinfo  {journal} {Annalen der Physik}\ }\textbf
  {\bibinfo {volume} {344}},\ \bibinfo {pages} {789} (\bibinfo {year}
  {1912})}\BibitemShut {NoStop}%
\bibitem [{\citenamefont {Huber}(2016)}]{Huber2016}%
  \BibitemOpen
  \bibfield  {author} {\bibinfo {author} {\bibfnamefont {S.~D.}\ \bibnamefont
  {Huber}},\ }\href {\doibase 10.1038/nphys3801} {\bibfield  {journal}
  {\bibinfo  {journal} {Nature Physics}\ }\textbf {\bibinfo {volume} {12}},\
  \bibinfo {pages} {621} (\bibinfo {year} {2016})}\BibitemShut {NoStop}%
\bibitem [{\citenamefont {Kane}\ and\ \citenamefont
  {Lubensky}(2014)}]{TM-Kane}%
  \BibitemOpen
  \bibfield  {author} {\bibinfo {author} {\bibfnamefont {C.~L.}\ \bibnamefont
  {Kane}}\ and\ \bibinfo {author} {\bibfnamefont {T.~C.}\ \bibnamefont
  {Lubensky}},\ }\href {\doibase 10.1038/nphys2835} {\bibfield  {journal}
  {\bibinfo  {journal} {Nature Physics}\ }\textbf {\bibinfo {volume} {10}},\
  \bibinfo {pages} {39} (\bibinfo {year} {2014})}\BibitemShut {NoStop}%
\bibitem [{\citenamefont {Mao}\ and\ \citenamefont
  {Lubensky}(2018)}]{TM-LubenskyRev}%
  \BibitemOpen
  \bibfield  {author} {\bibinfo {author} {\bibfnamefont {X.}~\bibnamefont
  {Mao}}\ and\ \bibinfo {author} {\bibfnamefont {T.~C.}\ \bibnamefont
  {Lubensky}},\ }\href {\doibase 10.1146/annurev-conmatphys-033117-054235}
  {\bibfield  {journal} {\bibinfo  {journal} {Annual Review of Condensed Matter
  Physics}\ }\textbf {\bibinfo {volume} {9}},\ \bibinfo {pages} {413} (\bibinfo
  {year} {2018})}\BibitemShut {NoStop}%
\bibitem [{\citenamefont {Ma}\ \emph {et~al.}(2019)\citenamefont {Ma},
  \citenamefont {Xiao},\ and\ \citenamefont {Chan}}]{Ma2019}%
  \BibitemOpen
  \bibfield  {author} {\bibinfo {author} {\bibfnamefont {G.}~\bibnamefont
  {Ma}}, \bibinfo {author} {\bibfnamefont {M.}~\bibnamefont {Xiao}}, \ and\
  \bibinfo {author} {\bibfnamefont {C.~T.}\ \bibnamefont {Chan}},\ }\href
  {\doibase 10.1038/s42254-019-0030-x} {\bibfield  {journal} {\bibinfo
  {journal} {Nature Reviews Physics}\ }\textbf {\bibinfo {volume} {1}},\
  \bibinfo {pages} {281} (\bibinfo {year} {2019})}\BibitemShut {NoStop}%
\bibitem [{\citenamefont {Süsstrunk}\ and\ \citenamefont
  {Huber}(2015)}]{HuberPendulum}%
  \BibitemOpen
  \bibfield  {author} {\bibinfo {author} {\bibfnamefont {R.}~\bibnamefont
  {Süsstrunk}}\ and\ \bibinfo {author} {\bibfnamefont {S.~D.}\ \bibnamefont
  {Huber}},\ }\href {\doibase 10.1126/science.aab0239} {\bibfield  {journal}
  {\bibinfo  {journal} {Science}\ }\textbf {\bibinfo {volume} {349}},\ \bibinfo
  {pages} {47} (\bibinfo {year} {2015})}\BibitemShut {NoStop}%
\bibitem [{\citenamefont {Nash}\ \emph {et~al.}(2015)\citenamefont {Nash},
  \citenamefont {Kleckner}, \citenamefont {Read}, \citenamefont {Vitelli},
  \citenamefont {Turner},\ and\ \citenamefont
  {Irvine}}]{GyroscopeMetamaterials}%
  \BibitemOpen
  \bibfield  {author} {\bibinfo {author} {\bibfnamefont {L.~M.}\ \bibnamefont
  {Nash}}, \bibinfo {author} {\bibfnamefont {D.}~\bibnamefont {Kleckner}},
  \bibinfo {author} {\bibfnamefont {A.}~\bibnamefont {Read}}, \bibinfo {author}
  {\bibfnamefont {V.}~\bibnamefont {Vitelli}}, \bibinfo {author} {\bibfnamefont
  {A.~M.}\ \bibnamefont {Turner}}, \ and\ \bibinfo {author} {\bibfnamefont
  {W.~T.~M.}\ \bibnamefont {Irvine}},\ }\href {\doibase
  10.1073/pnas.1507413112} {\bibfield  {journal} {\bibinfo  {journal}
  {Proceedings of the National Academy of Sciences}\ }\textbf {\bibinfo
  {volume} {112}},\ \bibinfo {pages} {14495} (\bibinfo {year}
  {2015})}\BibitemShut {NoStop}%
\bibitem [{\citenamefont {Apigo}\ \emph {et~al.}(2018)\citenamefont {Apigo},
  \citenamefont {Qian}, \citenamefont {Prodan},\ and\ \citenamefont
  {Prodan}}]{fidgetspinners}%
  \BibitemOpen
  \bibfield  {author} {\bibinfo {author} {\bibfnamefont {D.~J.}\ \bibnamefont
  {Apigo}}, \bibinfo {author} {\bibfnamefont {K.}~\bibnamefont {Qian}},
  \bibinfo {author} {\bibfnamefont {C.}~\bibnamefont {Prodan}}, \ and\ \bibinfo
  {author} {\bibfnamefont {E.}~\bibnamefont {Prodan}},\ }\href {\doibase
  10.1103/PhysRevMaterials.2.124203} {\bibfield  {journal} {\bibinfo  {journal}
  {Phys. Rev. Materials}\ }\textbf {\bibinfo {volume} {2}},\ \bibinfo {pages}
  {124203} (\bibinfo {year} {2018})}\BibitemShut {NoStop}%
\bibitem [{\citenamefont {Prodan}\ and\ \citenamefont
  {Prodan}(2009)}]{TM-Prodan}%
  \BibitemOpen
  \bibfield  {author} {\bibinfo {author} {\bibfnamefont {E.}~\bibnamefont
  {Prodan}}\ and\ \bibinfo {author} {\bibfnamefont {C.}~\bibnamefont
  {Prodan}},\ }\href {\doibase 10.1103/PhysRevLett.103.248101} {\bibfield
  {journal} {\bibinfo  {journal} {Phys. Rev. Lett.}\ }\textbf {\bibinfo
  {volume} {103}},\ \bibinfo {pages} {248101} (\bibinfo {year}
  {2009})}\BibitemShut {NoStop}%
\bibitem [{\citenamefont {S{\"u}sstrunk}\ and\ \citenamefont
  {Huber}(2016)}]{TM-Huber-classes}%
  \BibitemOpen
  \bibfield  {author} {\bibinfo {author} {\bibfnamefont {R.}~\bibnamefont
  {S{\"u}sstrunk}}\ and\ \bibinfo {author} {\bibfnamefont {S.~D.}\ \bibnamefont
  {Huber}},\ }\href {\doibase 10.1073/pnas.1605462113} {\bibfield  {journal}
  {\bibinfo  {journal} {Proceedings of the National Academy of Sciences}\
  }\textbf {\bibinfo {volume} {113}},\ \bibinfo {pages} {E4767} (\bibinfo
  {year} {2016})}\BibitemShut {NoStop}%
\bibitem [{\citenamefont {Peri}\ \emph {et~al.}(2020)\citenamefont {Peri},
  \citenamefont {Song}, \citenamefont {Serra-Garcia}, \citenamefont {Engeler},
  \citenamefont {Queiroz}, \citenamefont {Huang}, \citenamefont {Deng},
  \citenamefont {Liu}, \citenamefont {Bernevig},\ and\ \citenamefont
  {Huber}}]{TM-Huber-WeakTI}%
  \BibitemOpen
  \bibfield  {author} {\bibinfo {author} {\bibfnamefont {V.}~\bibnamefont
  {Peri}}, \bibinfo {author} {\bibfnamefont {Z.-D.}\ \bibnamefont {Song}},
  \bibinfo {author} {\bibfnamefont {M.}~\bibnamefont {Serra-Garcia}}, \bibinfo
  {author} {\bibfnamefont {P.}~\bibnamefont {Engeler}}, \bibinfo {author}
  {\bibfnamefont {R.}~\bibnamefont {Queiroz}}, \bibinfo {author} {\bibfnamefont
  {X.}~\bibnamefont {Huang}}, \bibinfo {author} {\bibfnamefont
  {W.}~\bibnamefont {Deng}}, \bibinfo {author} {\bibfnamefont {Z.}~\bibnamefont
  {Liu}}, \bibinfo {author} {\bibfnamefont {B.~A.}\ \bibnamefont {Bernevig}}, \
  and\ \bibinfo {author} {\bibfnamefont {S.~D.}\ \bibnamefont {Huber}},\ }\href
  {\doibase 10.1126/science.aaz7654} {\bibfield  {journal} {\bibinfo  {journal}
  {Science}\ }\textbf {\bibinfo {volume} {367}},\ \bibinfo {pages} {797}
  (\bibinfo {year} {2020})}\BibitemShut {NoStop}%
\bibitem [{\citenamefont {Deng}\ \emph {et~al.}(2020)\citenamefont {Deng},
  \citenamefont {Huang}, \citenamefont {Lu}, \citenamefont {Peri},
  \citenamefont {Li}, \citenamefont {Huber},\ and\ \citenamefont
  {Liu}}]{TM-Huber-SOC}%
  \BibitemOpen
  \bibfield  {author} {\bibinfo {author} {\bibfnamefont {W.}~\bibnamefont
  {Deng}}, \bibinfo {author} {\bibfnamefont {X.}~\bibnamefont {Huang}},
  \bibinfo {author} {\bibfnamefont {J.}~\bibnamefont {Lu}}, \bibinfo {author}
  {\bibfnamefont {V.}~\bibnamefont {Peri}}, \bibinfo {author} {\bibfnamefont
  {F.}~\bibnamefont {Li}}, \bibinfo {author} {\bibfnamefont {S.~D.}\
  \bibnamefont {Huber}}, \ and\ \bibinfo {author} {\bibfnamefont
  {Z.}~\bibnamefont {Liu}},\ }\href {\doibase 10.1038/s41467-020-17039-1}
  {\bibfield  {journal} {\bibinfo  {journal} {Nature Communications}\ }\textbf
  {\bibinfo {volume} {11}},\ \bibinfo {pages} {3227} (\bibinfo {year}
  {2020})}\BibitemShut {NoStop}%
\bibitem [{\citenamefont {Salerno}\ \emph {et~al.}(2017)\citenamefont
  {Salerno}, \citenamefont {Berardo}, \citenamefont {Ozawa}, \citenamefont
  {Price}, \citenamefont {Taxis}, \citenamefont {Pugno},\ and\ \citenamefont
  {Carusotto}}]{Salerno_2017NJP}%
  \BibitemOpen
  \bibfield  {author} {\bibinfo {author} {\bibfnamefont {G.}~\bibnamefont
  {Salerno}}, \bibinfo {author} {\bibfnamefont {A.}~\bibnamefont {Berardo}},
  \bibinfo {author} {\bibfnamefont {T.}~\bibnamefont {Ozawa}}, \bibinfo
  {author} {\bibfnamefont {H.~M.}\ \bibnamefont {Price}}, \bibinfo {author}
  {\bibfnamefont {L.}~\bibnamefont {Taxis}}, \bibinfo {author} {\bibfnamefont
  {N.~M.}\ \bibnamefont {Pugno}}, \ and\ \bibinfo {author} {\bibfnamefont
  {I.}~\bibnamefont {Carusotto}},\ }\href {\doibase 10.1088/1367-2630/aa6c03}
  {\bibfield  {journal} {\bibinfo  {journal} {New Journal of Physics}\ }\textbf
  {\bibinfo {volume} {19}},\ \bibinfo {pages} {055001} (\bibinfo {year}
  {2017})}\BibitemShut {NoStop}%
\bibitem [{\citenamefont {Cheng}\ \emph {et~al.}(2020)\citenamefont {Cheng},
  \citenamefont {Prodan},\ and\ \citenamefont {Prodan}}]{Camelia-DynPumping}%
  \BibitemOpen
  \bibfield  {author} {\bibinfo {author} {\bibfnamefont {W.}~\bibnamefont
  {Cheng}}, \bibinfo {author} {\bibfnamefont {E.}~\bibnamefont {Prodan}}, \
  and\ \bibinfo {author} {\bibfnamefont {C.}~\bibnamefont {Prodan}},\ }\href
  {\doibase 10.1103/PhysRevLett.125.224301} {\bibfield  {journal} {\bibinfo
  {journal} {Phys. Rev. Lett.}\ }\textbf {\bibinfo {volume} {125}},\ \bibinfo
  {pages} {224301} (\bibinfo {year} {2020})}\BibitemShut {NoStop}%
\bibitem [{\citenamefont {Qian}\ \emph {et~al.}(2020)\citenamefont {Qian},
  \citenamefont {Zhu}, \citenamefont {Ahn},\ and\ \citenamefont
  {Prodan}}]{Camelia-FlatBands}%
  \BibitemOpen
  \bibfield  {author} {\bibinfo {author} {\bibfnamefont {K.}~\bibnamefont
  {Qian}}, \bibinfo {author} {\bibfnamefont {L.}~\bibnamefont {Zhu}}, \bibinfo
  {author} {\bibfnamefont {K.~H.}\ \bibnamefont {Ahn}}, \ and\ \bibinfo
  {author} {\bibfnamefont {C.}~\bibnamefont {Prodan}},\ }\href {\doibase
  10.1103/PhysRevLett.125.225501} {\bibfield  {journal} {\bibinfo  {journal}
  {Phys. Rev. Lett.}\ }\textbf {\bibinfo {volume} {125}},\ \bibinfo {pages}
  {225501} (\bibinfo {year} {2020})}\BibitemShut {NoStop}%
\bibitem [{\citenamefont {Grinberg}\ \emph {et~al.}(2020)\citenamefont
  {Grinberg}, \citenamefont {Lin}, \citenamefont {Harris}, \citenamefont
  {Benalcazar}, \citenamefont {Peterson}, \citenamefont {Hughes},\ and\
  \citenamefont {Bahl}}]{TM-Bahl-Pumping}%
  \BibitemOpen
  \bibfield  {author} {\bibinfo {author} {\bibfnamefont {I.~H.}\ \bibnamefont
  {Grinberg}}, \bibinfo {author} {\bibfnamefont {M.}~\bibnamefont {Lin}},
  \bibinfo {author} {\bibfnamefont {C.}~\bibnamefont {Harris}}, \bibinfo
  {author} {\bibfnamefont {W.~A.}\ \bibnamefont {Benalcazar}}, \bibinfo
  {author} {\bibfnamefont {C.~W.}\ \bibnamefont {Peterson}}, \bibinfo {author}
  {\bibfnamefont {T.~L.}\ \bibnamefont {Hughes}}, \ and\ \bibinfo {author}
  {\bibfnamefont {G.}~\bibnamefont {Bahl}},\ }\href {\doibase
  10.1038/s41467-020-14804-0} {\bibfield  {journal} {\bibinfo  {journal}
  {Nature Communications}\ }\textbf {\bibinfo {volume} {11}},\ \bibinfo {pages}
  {974} (\bibinfo {year} {2020})}\BibitemShut {NoStop}%
\bibitem [{\citenamefont {Barlas}\ and\ \citenamefont
  {Prodan}(2018)}]{Prodan-passive}%
  \BibitemOpen
  \bibfield  {author} {\bibinfo {author} {\bibfnamefont {Y.}~\bibnamefont
  {Barlas}}\ and\ \bibinfo {author} {\bibfnamefont {E.}~\bibnamefont
  {Prodan}},\ }\href {\doibase 10.1103/PhysRevB.98.094310} {\bibfield
  {journal} {\bibinfo  {journal} {Phys. Rev. B}\ }\textbf {\bibinfo {volume}
  {98}},\ \bibinfo {pages} {094310} (\bibinfo {year} {2018})}\BibitemShut
  {NoStop}%
\bibitem [{\citenamefont {Serra-Garcia}\ \emph {et~al.}(2018)\citenamefont
  {Serra-Garcia}, \citenamefont {Peri}, \citenamefont {S{\"u}sstrunk},
  \citenamefont {Bilal}, \citenamefont {Larsen}, \citenamefont {Villanueva},\
  and\ \citenamefont {Huber}}]{Serra-Garcia2018}%
  \BibitemOpen
  \bibfield  {author} {\bibinfo {author} {\bibfnamefont {M.}~\bibnamefont
  {Serra-Garcia}}, \bibinfo {author} {\bibfnamefont {V.}~\bibnamefont {Peri}},
  \bibinfo {author} {\bibfnamefont {R.}~\bibnamefont {S{\"u}sstrunk}}, \bibinfo
  {author} {\bibfnamefont {O.~R.}\ \bibnamefont {Bilal}}, \bibinfo {author}
  {\bibfnamefont {T.}~\bibnamefont {Larsen}}, \bibinfo {author} {\bibfnamefont
  {L.~G.}\ \bibnamefont {Villanueva}}, \ and\ \bibinfo {author} {\bibfnamefont
  {S.~D.}\ \bibnamefont {Huber}},\ }\href {\doibase 10.1038/nature25156}
  {\bibfield  {journal} {\bibinfo  {journal} {Nature}\ }\textbf {\bibinfo
  {volume} {555}},\ \bibinfo {pages} {342} (\bibinfo {year}
  {2018})}\BibitemShut {NoStop}%
\bibitem [{\citenamefont {Salerno}\ \emph {et~al.}(2016)\citenamefont
  {Salerno}, \citenamefont {Ozawa}, \citenamefont {Price},\ and\ \citenamefont
  {Carusotto}}]{Salerno-alpha}%
  \BibitemOpen
  \bibfield  {author} {\bibinfo {author} {\bibfnamefont {G.}~\bibnamefont
  {Salerno}}, \bibinfo {author} {\bibfnamefont {T.}~\bibnamefont {Ozawa}},
  \bibinfo {author} {\bibfnamefont {H.~M.}\ \bibnamefont {Price}}, \ and\
  \bibinfo {author} {\bibfnamefont {I.}~\bibnamefont {Carusotto}},\ }\href
  {\doibase 10.1103/PhysRevB.93.085105} {\bibfield  {journal} {\bibinfo
  {journal} {Phys. Rev. B}\ }\textbf {\bibinfo {volume} {93}},\ \bibinfo
  {pages} {085105} (\bibinfo {year} {2016})}\BibitemShut {NoStop}%
\bibitem [{\citenamefont {Salerno}\ and\ \citenamefont
  {Carusotto}(2014)}]{Salerno_2014_epl}%
  \BibitemOpen
  \bibfield  {author} {\bibinfo {author} {\bibfnamefont {G.}~\bibnamefont
  {Salerno}}\ and\ \bibinfo {author} {\bibfnamefont {I.}~\bibnamefont
  {Carusotto}},\ }\href {\doibase 10.1209/0295-5075/106/24002} {\bibfield
  {journal} {\bibinfo  {journal} {{EPL} (Europhysics Letters)}\ }\textbf
  {\bibinfo {volume} {106}},\ \bibinfo {pages} {24002} (\bibinfo {year}
  {2014})}\BibitemShut {NoStop}%
\bibitem [{\citenamefont {Wang}\ \emph {et~al.}(2015)\citenamefont {Wang},
  \citenamefont {Luan},\ and\ \citenamefont {Zhang}}]{Wang_2015}%
  \BibitemOpen
  \bibfield  {author} {\bibinfo {author} {\bibfnamefont {Y.-T.}\ \bibnamefont
  {Wang}}, \bibinfo {author} {\bibfnamefont {P.-G.}\ \bibnamefont {Luan}}, \
  and\ \bibinfo {author} {\bibfnamefont {S.}~\bibnamefont {Zhang}},\ }\href
  {\doibase 10.1088/1367-2630/17/7/073031} {\bibfield  {journal} {\bibinfo
  {journal} {New Journal of Physics}\ }\textbf {\bibinfo {volume} {17}},\
  \bibinfo {pages} {073031} (\bibinfo {year} {2015})}\BibitemShut {NoStop}%
\bibitem [{\citenamefont {Mitchell}\ \emph {et~al.}(2018)\citenamefont
  {Mitchell}, \citenamefont {Nash}, \citenamefont {Hexner}, \citenamefont
  {Turner},\ and\ \citenamefont {Irvine}}]{Mitchell2018}%
  \BibitemOpen
  \bibfield  {author} {\bibinfo {author} {\bibfnamefont {N.~P.}\ \bibnamefont
  {Mitchell}}, \bibinfo {author} {\bibfnamefont {L.~M.}\ \bibnamefont {Nash}},
  \bibinfo {author} {\bibfnamefont {D.}~\bibnamefont {Hexner}}, \bibinfo
  {author} {\bibfnamefont {A.~M.}\ \bibnamefont {Turner}}, \ and\ \bibinfo
  {author} {\bibfnamefont {W.~T.~M.}\ \bibnamefont {Irvine}},\ }\href {\doibase
  10.1038/s41567-017-0024-5} {\bibfield  {journal} {\bibinfo  {journal} {Nature
  Physics}\ }\textbf {\bibinfo {volume} {14}},\ \bibinfo {pages} {380}
  (\bibinfo {year} {2018})}\BibitemShut {NoStop}%
\bibitem [{\citenamefont {Sirota}\ \emph {et~al.}(2020)\citenamefont {Sirota},
  \citenamefont {Ilan}, \citenamefont {Shokef},\ and\ \citenamefont
  {Lahini}}]{Ilan-prop}%
  \BibitemOpen
  \bibfield  {author} {\bibinfo {author} {\bibfnamefont {L.}~\bibnamefont
  {Sirota}}, \bibinfo {author} {\bibfnamefont {R.}~\bibnamefont {Ilan}},
  \bibinfo {author} {\bibfnamefont {Y.}~\bibnamefont {Shokef}}, \ and\ \bibinfo
  {author} {\bibfnamefont {Y.}~\bibnamefont {Lahini}},\ }\href {\doibase
  10.1103/PhysRevLett.125.256802} {\bibfield  {journal} {\bibinfo  {journal}
  {Phys. Rev. Lett.}\ }\textbf {\bibinfo {volume} {125}},\ \bibinfo {pages}
  {256802} (\bibinfo {year} {2020})}\BibitemShut {NoStop}%
\bibitem [{\citenamefont {Brandenbourger}\ \emph {et~al.}(2019)\citenamefont
  {Brandenbourger}, \citenamefont {Locsin}, \citenamefont {Lerner},\ and\
  \citenamefont {Coulais}}]{Brandenbourger2019}%
  \BibitemOpen
  \bibfield  {author} {\bibinfo {author} {\bibfnamefont {M.}~\bibnamefont
  {Brandenbourger}}, \bibinfo {author} {\bibfnamefont {X.}~\bibnamefont
  {Locsin}}, \bibinfo {author} {\bibfnamefont {E.}~\bibnamefont {Lerner}}, \
  and\ \bibinfo {author} {\bibfnamefont {C.}~\bibnamefont {Coulais}},\ }\href
  {\doibase 10.1038/s41467-019-12599-3} {\bibfield  {journal} {\bibinfo
  {journal} {Nature Communications}\ }\textbf {\bibinfo {volume} {10}},\
  \bibinfo {pages} {4608} (\bibinfo {year} {2019})}\BibitemShut {NoStop}%
\bibitem [{\citenamefont {Ghatak}\ \emph {et~al.}(2020)\citenamefont {Ghatak},
  \citenamefont {Brandenbourger}, \citenamefont {van Wezel},\ and\
  \citenamefont {Coulais}}]{Coulais-PNAS}%
  \BibitemOpen
  \bibfield  {author} {\bibinfo {author} {\bibfnamefont {A.}~\bibnamefont
  {Ghatak}}, \bibinfo {author} {\bibfnamefont {M.}~\bibnamefont
  {Brandenbourger}}, \bibinfo {author} {\bibfnamefont {J.}~\bibnamefont {van
  Wezel}}, \ and\ \bibinfo {author} {\bibfnamefont {C.}~\bibnamefont
  {Coulais}},\ }\href {\doibase 10.1073/pnas.2010580117} {\bibfield  {journal}
  {\bibinfo  {journal} {Proceedings of the National Academy of Sciences}\ }
  (\bibinfo {year} {2020}),\ 10.1073/pnas.2010580117}\BibitemShut {NoStop}%
\bibitem [{Note1()}]{Note1}%
  \BibitemOpen
  \bibinfo {note} {Here we consider only instantaneous dependencies, but this
  approach also allows for time-retarded interactions}\BibitemShut {NoStop}%
\bibitem [{\citenamefont {Pouladian-Kari}(1990)}]{SolenoidDesign}%
  \BibitemOpen
  \bibfield  {author} {\bibinfo {author} {\bibfnamefont {R.}~\bibnamefont
  {Pouladian-Kari}},\ }\href {\doibase 10.1088/0957-0233/1/12/023} {\bibfield
  {journal} {\bibinfo  {journal} {Measurement Science and Technology}\ }\textbf
  {\bibinfo {volume} {1}},\ \bibinfo {pages} {1377} (\bibinfo {year}
  {1990})}\BibitemShut {NoStop}%
\bibitem [{\citenamefont {Fruchart}\ \emph {et~al.}(2021)\citenamefont
  {Fruchart}, \citenamefont {Hanai}, \citenamefont {Littlewood},\ and\
  \citenamefont {Vitelli}}]{Vitelli-NonRecip}%
  \BibitemOpen
  \bibfield  {author} {\bibinfo {author} {\bibfnamefont {M.}~\bibnamefont
  {Fruchart}}, \bibinfo {author} {\bibfnamefont {R.}~\bibnamefont {Hanai}},
  \bibinfo {author} {\bibfnamefont {P.~B.}\ \bibnamefont {Littlewood}}, \ and\
  \bibinfo {author} {\bibfnamefont {V.}~\bibnamefont {Vitelli}},\ }\href
  {\doibase 10.1038/s41586-021-03375-9} {\bibfield  {journal} {\bibinfo
  {journal} {Nature}\ }\textbf {\bibinfo {volume} {592}},\ \bibinfo {pages}
  {363} (\bibinfo {year} {2021})}\BibitemShut {NoStop}%
\bibitem [{\citenamefont {Ashida}\ \emph {et~al.}(2020)\citenamefont {Ashida},
  \citenamefont {Gong},\ and\ \citenamefont {Ueda}}]{Ueda-NH-review}%
  \BibitemOpen
  \bibfield  {author} {\bibinfo {author} {\bibfnamefont {Y.}~\bibnamefont
  {Ashida}}, \bibinfo {author} {\bibfnamefont {Z.}~\bibnamefont {Gong}}, \ and\
  \bibinfo {author} {\bibfnamefont {M.}~\bibnamefont {Ueda}},\ }\href {\doibase
  10.1080/00018732.2021.1876991} {\bibfield  {journal} {\bibinfo  {journal}
  {Advances in Physics}\ }\textbf {\bibinfo {volume} {69}},\ \bibinfo {pages}
  {249} (\bibinfo {year} {2020})}\BibitemShut {NoStop}%
\bibitem [{\citenamefont {Shankar}\ \emph {et~al.}(2020)\citenamefont
  {Shankar}, \citenamefont {Souslov}, \citenamefont {Bowick}, \citenamefont
  {Marchetti},\ and\ \citenamefont {Vitelli}}]{shankar2020topological}%
  \BibitemOpen
  \bibfield  {author} {\bibinfo {author} {\bibfnamefont {S.}~\bibnamefont
  {Shankar}}, \bibinfo {author} {\bibfnamefont {A.}~\bibnamefont {Souslov}},
  \bibinfo {author} {\bibfnamefont {M.~J.}\ \bibnamefont {Bowick}}, \bibinfo
  {author} {\bibfnamefont {M.~C.}\ \bibnamefont {Marchetti}}, \ and\ \bibinfo
  {author} {\bibfnamefont {V.}~\bibnamefont {Vitelli}},\ }\href@noop {}
  {\enquote {\bibinfo {title} {Topological active matter},}\ } (\bibinfo {year}
  {2020}),\ \Eprint {http://arxiv.org/abs/2010.00364} {arXiv:2010.00364
  [cond-mat.soft]} \BibitemShut {NoStop}%
\bibitem [{\citenamefont {Gou}\ \emph {et~al.}(2020)\citenamefont {Gou},
  \citenamefont {Chen}, \citenamefont {Xie}, \citenamefont {Xiao},
  \citenamefont {Deng}, \citenamefont {Gadway}, \citenamefont {Yi},\ and\
  \citenamefont {Yan}}]{Bo-NonRecip}%
  \BibitemOpen
  \bibfield  {author} {\bibinfo {author} {\bibfnamefont {W.}~\bibnamefont
  {Gou}}, \bibinfo {author} {\bibfnamefont {T.}~\bibnamefont {Chen}}, \bibinfo
  {author} {\bibfnamefont {D.}~\bibnamefont {Xie}}, \bibinfo {author}
  {\bibfnamefont {T.}~\bibnamefont {Xiao}}, \bibinfo {author} {\bibfnamefont
  {T.-S.}\ \bibnamefont {Deng}}, \bibinfo {author} {\bibfnamefont
  {B.}~\bibnamefont {Gadway}}, \bibinfo {author} {\bibfnamefont
  {W.}~\bibnamefont {Yi}}, \ and\ \bibinfo {author} {\bibfnamefont
  {B.}~\bibnamefont {Yan}},\ }\href {\doibase 10.1103/PhysRevLett.124.070402}
  {\bibfield  {journal} {\bibinfo  {journal} {Phys. Rev. Lett.}\ }\textbf
  {\bibinfo {volume} {124}},\ \bibinfo {pages} {070402} (\bibinfo {year}
  {2020})}\BibitemShut {NoStop}%
\bibitem [{\citenamefont {Hatano}\ and\ \citenamefont
  {Nelson}(1996)}]{HN-PRL-96}%
  \BibitemOpen
  \bibfield  {author} {\bibinfo {author} {\bibfnamefont {N.}~\bibnamefont
  {Hatano}}\ and\ \bibinfo {author} {\bibfnamefont {D.~R.}\ \bibnamefont
  {Nelson}},\ }\href {\doibase 10.1103/PhysRevLett.77.570} {\bibfield
  {journal} {\bibinfo  {journal} {Phys. Rev. Lett.}\ }\textbf {\bibinfo
  {volume} {77}},\ \bibinfo {pages} {570} (\bibinfo {year} {1996})}\BibitemShut
  {NoStop}%
\bibitem [{\citenamefont {Hatano}\ and\ \citenamefont
  {Nelson}(1997)}]{HN-PRB-97}%
  \BibitemOpen
  \bibfield  {author} {\bibinfo {author} {\bibfnamefont {N.}~\bibnamefont
  {Hatano}}\ and\ \bibinfo {author} {\bibfnamefont {D.~R.}\ \bibnamefont
  {Nelson}},\ }\href {\doibase 10.1103/PhysRevB.56.8651} {\bibfield  {journal}
  {\bibinfo  {journal} {Phys. Rev. B}\ }\textbf {\bibinfo {volume} {56}},\
  \bibinfo {pages} {8651} (\bibinfo {year} {1997})}\BibitemShut {NoStop}%
\bibitem [{\citenamefont {Hatano}\ and\ \citenamefont
  {Nelson}(1998)}]{HN-PRB-98}%
  \BibitemOpen
  \bibfield  {author} {\bibinfo {author} {\bibfnamefont {N.}~\bibnamefont
  {Hatano}}\ and\ \bibinfo {author} {\bibfnamefont {D.~R.}\ \bibnamefont
  {Nelson}},\ }\href {\doibase 10.1103/PhysRevB.58.8384} {\bibfield  {journal}
  {\bibinfo  {journal} {Phys. Rev. B}\ }\textbf {\bibinfo {volume} {58}},\
  \bibinfo {pages} {8384} (\bibinfo {year} {1998})}\BibitemShut {NoStop}%
\bibitem [{\citenamefont {Liu}\ \emph {et~al.}(2021)\citenamefont {Liu},
  \citenamefont {Shao}, \citenamefont {Ma}, \citenamefont {Zhang},
  \citenamefont {You}, \citenamefont {Wu}, \citenamefont {Xiang}, \citenamefont
  {Cui},\ and\ \citenamefont {Zhang}}]{Non-Herm-Skin}%
  \BibitemOpen
  \bibfield  {author} {\bibinfo {author} {\bibfnamefont {S.}~\bibnamefont
  {Liu}}, \bibinfo {author} {\bibfnamefont {R.}~\bibnamefont {Shao}}, \bibinfo
  {author} {\bibfnamefont {S.}~\bibnamefont {Ma}}, \bibinfo {author}
  {\bibfnamefont {L.}~\bibnamefont {Zhang}}, \bibinfo {author} {\bibfnamefont
  {O.}~\bibnamefont {You}}, \bibinfo {author} {\bibfnamefont {H.}~\bibnamefont
  {Wu}}, \bibinfo {author} {\bibfnamefont {Y.~J.}\ \bibnamefont {Xiang}},
  \bibinfo {author} {\bibfnamefont {T.~J.}\ \bibnamefont {Cui}}, \ and\
  \bibinfo {author} {\bibfnamefont {S.}~\bibnamefont {Zhang}},\ }\href
  {\doibase 10.34133/2021/5608038} {\bibfield  {journal} {\bibinfo  {journal}
  {Research}\ }\textbf {\bibinfo {volume} {2021}},\ \bibinfo {pages} {5608038}
  (\bibinfo {year} {2021})}\BibitemShut {NoStop}%
\bibitem [{\citenamefont {Duncan}\ \emph {et~al.}(2020)\citenamefont {Duncan},
  \citenamefont {Hartmann}, \citenamefont {Thomson},\ and\ \citenamefont
  {{\"O}hberg}}]{Duncan2020}%
  \BibitemOpen
  \bibfield  {author} {\bibinfo {author} {\bibfnamefont {C.~W.}\ \bibnamefont
  {Duncan}}, \bibinfo {author} {\bibfnamefont {M.~J.}\ \bibnamefont
  {Hartmann}}, \bibinfo {author} {\bibfnamefont {R.~R.}\ \bibnamefont
  {Thomson}}, \ and\ \bibinfo {author} {\bibfnamefont {P.}~\bibnamefont
  {{\"O}hberg}},\ }\href {\doibase 10.1140/epjd/e2020-100521-0} {\bibfield
  {journal} {\bibinfo  {journal} {The European Physical Journal D}\ }\textbf
  {\bibinfo {volume} {74}},\ \bibinfo {pages} {84} (\bibinfo {year}
  {2020})}\BibitemShut {NoStop}%
\bibitem [{\citenamefont {Hadad}\ \emph {et~al.}(2018)\citenamefont {Hadad},
  \citenamefont {Soric}, \citenamefont {Khanikaev},\ and\ \citenamefont
  {Al{\`u}}}]{Khanikaev-nonlinear}%
  \BibitemOpen
  \bibfield  {author} {\bibinfo {author} {\bibfnamefont {Y.}~\bibnamefont
  {Hadad}}, \bibinfo {author} {\bibfnamefont {J.~C.}\ \bibnamefont {Soric}},
  \bibinfo {author} {\bibfnamefont {A.~B.}\ \bibnamefont {Khanikaev}}, \ and\
  \bibinfo {author} {\bibfnamefont {A.}~\bibnamefont {Al{\`u}}},\ }\href
  {\doibase 10.1038/s41928-018-0042-z} {\bibfield  {journal} {\bibinfo
  {journal} {Nature Electronics}\ }\textbf {\bibinfo {volume} {1}},\ \bibinfo
  {pages} {178} (\bibinfo {year} {2018})}\BibitemShut {NoStop}%
\bibitem [{\citenamefont {Raghavan}\ \emph {et~al.}(1999)\citenamefont
  {Raghavan}, \citenamefont {Smerzi}, \citenamefont {Fantoni},\ and\
  \citenamefont {Shenoy}}]{Raghavan}%
  \BibitemOpen
  \bibfield  {author} {\bibinfo {author} {\bibfnamefont {S.}~\bibnamefont
  {Raghavan}}, \bibinfo {author} {\bibfnamefont {A.}~\bibnamefont {Smerzi}},
  \bibinfo {author} {\bibfnamefont {S.}~\bibnamefont {Fantoni}}, \ and\
  \bibinfo {author} {\bibfnamefont {S.~R.}\ \bibnamefont {Shenoy}},\ }\href
  {\doibase 10.1103/PhysRevA.59.620} {\bibfield  {journal} {\bibinfo  {journal}
  {Phys. Rev. A}\ }\textbf {\bibinfo {volume} {59}},\ \bibinfo {pages} {620}
  (\bibinfo {year} {1999})}\BibitemShut {NoStop}%
\bibitem [{\citenamefont {Juli\'a-D\'{\i}az}\ \emph {et~al.}(2010)\citenamefont
  {Juli\'a-D\'{\i}az}, \citenamefont {Dagnino}, \citenamefont {Lewenstein},
  \citenamefont {Martorell},\ and\ \citenamefont {Polls}}]{Maciej-self-trap}%
  \BibitemOpen
  \bibfield  {author} {\bibinfo {author} {\bibfnamefont {B.}~\bibnamefont
  {Juli\'a-D\'{\i}az}}, \bibinfo {author} {\bibfnamefont {D.}~\bibnamefont
  {Dagnino}}, \bibinfo {author} {\bibfnamefont {M.}~\bibnamefont {Lewenstein}},
  \bibinfo {author} {\bibfnamefont {J.}~\bibnamefont {Martorell}}, \ and\
  \bibinfo {author} {\bibfnamefont {A.}~\bibnamefont {Polls}},\ }\href
  {\doibase 10.1103/PhysRevA.81.023615} {\bibfield  {journal} {\bibinfo
  {journal} {Phys. Rev. A}\ }\textbf {\bibinfo {volume} {81}},\ \bibinfo
  {pages} {023615} (\bibinfo {year} {2010})}\BibitemShut {NoStop}%
\bibitem [{\citenamefont {Abbarchi}\ \emph {et~al.}(2013)\citenamefont
  {Abbarchi}, \citenamefont {Amo}, \citenamefont {Sala}, \citenamefont
  {Solnyshkov}, \citenamefont {Flayac}, \citenamefont {Ferrier}, \citenamefont
  {Sagnes}, \citenamefont {Galopin}, \citenamefont {Lema{\^i}tre},
  \citenamefont {Malpuech},\ and\ \citenamefont {Bloch}}]{Bloch-polariton}%
  \BibitemOpen
  \bibfield  {author} {\bibinfo {author} {\bibfnamefont {M.}~\bibnamefont
  {Abbarchi}}, \bibinfo {author} {\bibfnamefont {A.}~\bibnamefont {Amo}},
  \bibinfo {author} {\bibfnamefont {V.~G.}\ \bibnamefont {Sala}}, \bibinfo
  {author} {\bibfnamefont {D.~D.}\ \bibnamefont {Solnyshkov}}, \bibinfo
  {author} {\bibfnamefont {H.}~\bibnamefont {Flayac}}, \bibinfo {author}
  {\bibfnamefont {L.}~\bibnamefont {Ferrier}}, \bibinfo {author} {\bibfnamefont
  {I.}~\bibnamefont {Sagnes}}, \bibinfo {author} {\bibfnamefont
  {E.}~\bibnamefont {Galopin}}, \bibinfo {author} {\bibfnamefont
  {A.}~\bibnamefont {Lema{\^i}tre}}, \bibinfo {author} {\bibfnamefont
  {G.}~\bibnamefont {Malpuech}}, \ and\ \bibinfo {author} {\bibfnamefont
  {J.}~\bibnamefont {Bloch}},\ }\href {\doibase 10.1038/nphys2609} {\bibfield
  {journal} {\bibinfo  {journal} {Nature Physics}\ }\textbf {\bibinfo {volume}
  {9}},\ \bibinfo {pages} {275} (\bibinfo {year} {2013})}\BibitemShut {NoStop}%
\bibitem [{\citenamefont {Albiez}\ \emph {et~al.}(2005)\citenamefont {Albiez},
  \citenamefont {Gati}, \citenamefont {F\"olling}, \citenamefont {Hunsmann},
  \citenamefont {Cristiani},\ and\ \citenamefont
  {Oberthaler}}]{Albiez-DirectJosephson}%
  \BibitemOpen
  \bibfield  {author} {\bibinfo {author} {\bibfnamefont {M.}~\bibnamefont
  {Albiez}}, \bibinfo {author} {\bibfnamefont {R.}~\bibnamefont {Gati}},
  \bibinfo {author} {\bibfnamefont {J.}~\bibnamefont {F\"olling}}, \bibinfo
  {author} {\bibfnamefont {S.}~\bibnamefont {Hunsmann}}, \bibinfo {author}
  {\bibfnamefont {M.}~\bibnamefont {Cristiani}}, \ and\ \bibinfo {author}
  {\bibfnamefont {M.~K.}\ \bibnamefont {Oberthaler}},\ }\href {\doibase
  10.1103/PhysRevLett.95.010402} {\bibfield  {journal} {\bibinfo  {journal}
  {Phys. Rev. Lett.}\ }\textbf {\bibinfo {volume} {95}},\ \bibinfo {pages}
  {010402} (\bibinfo {year} {2005})}\BibitemShut {NoStop}%
\bibitem [{\citenamefont {Levy}\ \emph {et~al.}(2007)\citenamefont {Levy},
  \citenamefont {Lahoud}, \citenamefont {Shomroni},\ and\ \citenamefont
  {Steinhauer}}]{Levy2007}%
  \BibitemOpen
  \bibfield  {author} {\bibinfo {author} {\bibfnamefont {S.}~\bibnamefont
  {Levy}}, \bibinfo {author} {\bibfnamefont {E.}~\bibnamefont {Lahoud}},
  \bibinfo {author} {\bibfnamefont {I.}~\bibnamefont {Shomroni}}, \ and\
  \bibinfo {author} {\bibfnamefont {J.}~\bibnamefont {Steinhauer}},\ }\href
  {\doibase 10.1038/nature06186} {\bibfield  {journal} {\bibinfo  {journal}
  {Nature}\ }\textbf {\bibinfo {volume} {449}},\ \bibinfo {pages} {579}
  (\bibinfo {year} {2007})}\BibitemShut {NoStop}%
\bibitem [{\citenamefont {Anker}\ \emph {et~al.}(2005)\citenamefont {Anker},
  \citenamefont {Albiez}, \citenamefont {Gati}, \citenamefont {Hunsmann},
  \citenamefont {Eiermann}, \citenamefont {Trombettoni},\ and\ \citenamefont
  {Oberthaler}}]{Anker-SelfTrap}%
  \BibitemOpen
  \bibfield  {author} {\bibinfo {author} {\bibfnamefont {T.}~\bibnamefont
  {Anker}}, \bibinfo {author} {\bibfnamefont {M.}~\bibnamefont {Albiez}},
  \bibinfo {author} {\bibfnamefont {R.}~\bibnamefont {Gati}}, \bibinfo {author}
  {\bibfnamefont {S.}~\bibnamefont {Hunsmann}}, \bibinfo {author}
  {\bibfnamefont {B.}~\bibnamefont {Eiermann}}, \bibinfo {author}
  {\bibfnamefont {A.}~\bibnamefont {Trombettoni}}, \ and\ \bibinfo {author}
  {\bibfnamefont {M.~K.}\ \bibnamefont {Oberthaler}},\ }\href {\doibase
  10.1103/PhysRevLett.94.020403} {\bibfield  {journal} {\bibinfo  {journal}
  {Phys. Rev. Lett.}\ }\textbf {\bibinfo {volume} {94}},\ \bibinfo {pages}
  {020403} (\bibinfo {year} {2005})}\BibitemShut {NoStop}%
\bibitem [{\citenamefont {Zibold}\ \emph {et~al.}(2010)\citenamefont {Zibold},
  \citenamefont {Nicklas}, \citenamefont {Gross},\ and\ \citenamefont
  {Oberthaler}}]{Zibold}%
  \BibitemOpen
  \bibfield  {author} {\bibinfo {author} {\bibfnamefont {T.}~\bibnamefont
  {Zibold}}, \bibinfo {author} {\bibfnamefont {E.}~\bibnamefont {Nicklas}},
  \bibinfo {author} {\bibfnamefont {C.}~\bibnamefont {Gross}}, \ and\ \bibinfo
  {author} {\bibfnamefont {M.~K.}\ \bibnamefont {Oberthaler}},\ }\href
  {\doibase 10.1103/PhysRevLett.105.204101} {\bibfield  {journal} {\bibinfo
  {journal} {Phys. Rev. Lett.}\ }\textbf {\bibinfo {volume} {105}},\ \bibinfo
  {pages} {204101} (\bibinfo {year} {2010})}\BibitemShut {NoStop}%
\bibitem [{\citenamefont {Chiao}\ \emph {et~al.}(1964)\citenamefont {Chiao},
  \citenamefont {Garmire},\ and\ \citenamefont {Townes}}]{chiao-selftrap}%
  \BibitemOpen
  \bibfield  {author} {\bibinfo {author} {\bibfnamefont {R.~Y.}\ \bibnamefont
  {Chiao}}, \bibinfo {author} {\bibfnamefont {E.}~\bibnamefont {Garmire}}, \
  and\ \bibinfo {author} {\bibfnamefont {C.~H.}\ \bibnamefont {Townes}},\
  }\href {\doibase 10.1103/PhysRevLett.13.479} {\bibfield  {journal} {\bibinfo
  {journal} {Phys. Rev. Lett.}\ }\textbf {\bibinfo {volume} {13}},\ \bibinfo
  {pages} {479} (\bibinfo {year} {1964})}\BibitemShut {NoStop}%
\bibitem [{\citenamefont {Hasegawa}\ and\ \citenamefont
  {Tappert}(1973)}]{tappert}%
  \BibitemOpen
  \bibfield  {author} {\bibinfo {author} {\bibfnamefont {A.}~\bibnamefont
  {Hasegawa}}\ and\ \bibinfo {author} {\bibfnamefont {F.}~\bibnamefont
  {Tappert}},\ }\href {\doibase 10.1063/1.1654836} {\bibfield  {journal}
  {\bibinfo  {journal} {Applied Physics Letters}\ }\textbf {\bibinfo {volume}
  {23}},\ \bibinfo {pages} {142} (\bibinfo {year} {1973})}\BibitemShut
  {NoStop}%
\bibitem [{\citenamefont {Mollenauer}\ \emph {et~al.}(1980)\citenamefont
  {Mollenauer}, \citenamefont {Stolen},\ and\ \citenamefont {Gordon}}]{gordon}%
  \BibitemOpen
  \bibfield  {author} {\bibinfo {author} {\bibfnamefont {L.~F.}\ \bibnamefont
  {Mollenauer}}, \bibinfo {author} {\bibfnamefont {R.~H.}\ \bibnamefont
  {Stolen}}, \ and\ \bibinfo {author} {\bibfnamefont {J.~P.}\ \bibnamefont
  {Gordon}},\ }\href {\doibase 10.1103/PhysRevLett.45.1095} {\bibfield
  {journal} {\bibinfo  {journal} {Phys. Rev. Lett.}\ }\textbf {\bibinfo
  {volume} {45}},\ \bibinfo {pages} {1095} (\bibinfo {year}
  {1980})}\BibitemShut {NoStop}%
\bibitem [{\citenamefont {Snyder}\ \emph {et~al.}(1991)\citenamefont {Snyder},
  \citenamefont {Mitchell}, \citenamefont {Poladian},\ and\ \citenamefont
  {Ladouceur}}]{Snyder:91}%
  \BibitemOpen
  \bibfield  {author} {\bibinfo {author} {\bibfnamefont {A.~W.}\ \bibnamefont
  {Snyder}}, \bibinfo {author} {\bibfnamefont {D.~J.}\ \bibnamefont
  {Mitchell}}, \bibinfo {author} {\bibfnamefont {L.}~\bibnamefont {Poladian}},
  \ and\ \bibinfo {author} {\bibfnamefont {F.}~\bibnamefont {Ladouceur}},\
  }\href {\doibase 10.1364/OL.16.000021} {\bibfield  {journal} {\bibinfo
  {journal} {Opt. Lett.}\ }\textbf {\bibinfo {volume} {16}},\ \bibinfo {pages}
  {21} (\bibinfo {year} {1991})}\BibitemShut {NoStop}%
\bibitem [{Note2()}]{Note2}%
  \BibitemOpen
  \bibinfo {note} {However, the roles of the two modes are reversed, as we
  operate with a negative $J$ value (effectively a $\pi $ tunneling phase). In
  the canonical bosonic Josephson junction~\cite {Raghavan}, the out-of-phase
  $\pi $-mode undergoes mode-softening while the in-phase plasma mode is purely
  stiffened.}\BibitemShut {Stop}%
\bibitem [{\citenamefont {Bukov}\ \emph {et~al.}(2015)\citenamefont {Bukov},
  \citenamefont {D'Alessio},\ and\ \citenamefont
  {Polkovnikov}}]{floquet-bukov}%
  \BibitemOpen
  \bibfield  {author} {\bibinfo {author} {\bibfnamefont {M.}~\bibnamefont
  {Bukov}}, \bibinfo {author} {\bibfnamefont {L.}~\bibnamefont {D'Alessio}}, \
  and\ \bibinfo {author} {\bibfnamefont {A.}~\bibnamefont {Polkovnikov}},\
  }\href {\doibase 10.1080/00018732.2015.1055918} {\bibfield  {journal}
  {\bibinfo  {journal} {Advances in Physics}\ }\textbf {\bibinfo {volume}
  {64}},\ \bibinfo {pages} {139} (\bibinfo {year} {2015})}\BibitemShut
  {NoStop}%
\bibitem [{\citenamefont {Casati}\ and\ \citenamefont
  {Molinari}(1989)}]{casati}%
  \BibitemOpen
  \bibfield  {author} {\bibinfo {author} {\bibfnamefont {G.}~\bibnamefont
  {Casati}}\ and\ \bibinfo {author} {\bibfnamefont {L.}~\bibnamefont
  {Molinari}},\ }\href {\doibase 10.1143/PTPS.98.287} {\bibfield  {journal}
  {\bibinfo  {journal} {Progress of Theoretical Physics Supplement}\ }\textbf
  {\bibinfo {volume} {98}},\ \bibinfo {pages} {287} (\bibinfo {year}
  {1989})}\BibitemShut {NoStop}%
\bibitem [{\citenamefont {Matus}\ \emph {et~al.}(2021)\citenamefont {Matus},
  \citenamefont {Giergiel},\ and\ \citenamefont {Sacha}}]{Anderson-molecule}%
  \BibitemOpen
  \bibfield  {author} {\bibinfo {author} {\bibfnamefont {P.}~\bibnamefont
  {Matus}}, \bibinfo {author} {\bibfnamefont {K.}~\bibnamefont {Giergiel}}, \
  and\ \bibinfo {author} {\bibfnamefont {K.}~\bibnamefont {Sacha}},\ }\href
  {\doibase 10.1103/PhysRevA.103.023320} {\bibfield  {journal} {\bibinfo
  {journal} {Phys. Rev. A}\ }\textbf {\bibinfo {volume} {103}},\ \bibinfo
  {pages} {023320} (\bibinfo {year} {2021})}\BibitemShut {NoStop}%
\bibitem [{\citenamefont {Pretko}(2019)}]{pretko-circuit}%
  \BibitemOpen
  \bibfield  {author} {\bibinfo {author} {\bibfnamefont {M.}~\bibnamefont
  {Pretko}},\ }\href {\doibase 10.1103/PhysRevB.100.245103} {\bibfield
  {journal} {\bibinfo  {journal} {Phys. Rev. B}\ }\textbf {\bibinfo {volume}
  {100}},\ \bibinfo {pages} {245103} (\bibinfo {year} {2019})}\BibitemShut
  {NoStop}%
\bibitem [{\citenamefont {Mil}\ \emph {et~al.}(2019)\citenamefont {Mil},
  \citenamefont {Zache}, \citenamefont {Hegde}, \citenamefont {Xia},
  \citenamefont {Bhatt}, \citenamefont {Oberthaler}, \citenamefont {Hauke},
  \citenamefont {Berges},\ and\ \citenamefont {Jendrzejewski}}]{Jendrz-2019}%
  \BibitemOpen
  \bibfield  {author} {\bibinfo {author} {\bibfnamefont {A.}~\bibnamefont
  {Mil}}, \bibinfo {author} {\bibfnamefont {T.~V.}\ \bibnamefont {Zache}},
  \bibinfo {author} {\bibfnamefont {A.}~\bibnamefont {Hegde}}, \bibinfo
  {author} {\bibfnamefont {A.}~\bibnamefont {Xia}}, \bibinfo {author}
  {\bibfnamefont {R.~P.}\ \bibnamefont {Bhatt}}, \bibinfo {author}
  {\bibfnamefont {M.~K.}\ \bibnamefont {Oberthaler}}, \bibinfo {author}
  {\bibfnamefont {P.}~\bibnamefont {Hauke}}, \bibinfo {author} {\bibfnamefont
  {J.}~\bibnamefont {Berges}}, \ and\ \bibinfo {author} {\bibfnamefont
  {F.}~\bibnamefont {Jendrzejewski}},\ }\href@noop {} {\bibfield  {journal}
  {\bibinfo  {journal} {arXiv:1909.07641}\ } (\bibinfo {year}
  {2019})}\BibitemShut {NoStop}%
\bibitem [{\citenamefont {Petiziol}\ \emph {et~al.}(2020)\citenamefont
  {Petiziol}, \citenamefont {Sameti}, \citenamefont {Carretta}, \citenamefont
  {Wimberger},\ and\ \citenamefont {Mintert}}]{mintert-3body}%
  \BibitemOpen
  \bibfield  {author} {\bibinfo {author} {\bibfnamefont {F.}~\bibnamefont
  {Petiziol}}, \bibinfo {author} {\bibfnamefont {M.}~\bibnamefont {Sameti}},
  \bibinfo {author} {\bibfnamefont {S.}~\bibnamefont {Carretta}}, \bibinfo
  {author} {\bibfnamefont {S.}~\bibnamefont {Wimberger}}, \ and\ \bibinfo
  {author} {\bibfnamefont {F.}~\bibnamefont {Mintert}},\ }\href@noop {}
  {\enquote {\bibinfo {title} {Quantum simulation of three-body interactions in
  weakly driven quantum systems},}\ } (\bibinfo {year} {2020}),\ \Eprint
  {http://arxiv.org/abs/2011.03399} {arXiv:2011.03399 [quant-ph]} \BibitemShut
  {NoStop}%
\bibitem [{\citenamefont {Lapa}\ \emph {et~al.}(2016)\citenamefont {Lapa},
  \citenamefont {Teo},\ and\ \citenamefont {Hughes}}]{lapa-int-enab}%
  \BibitemOpen
  \bibfield  {author} {\bibinfo {author} {\bibfnamefont {M.~F.}\ \bibnamefont
  {Lapa}}, \bibinfo {author} {\bibfnamefont {J.~C.~Y.}\ \bibnamefont {Teo}}, \
  and\ \bibinfo {author} {\bibfnamefont {T.~L.}\ \bibnamefont {Hughes}},\
  }\href {\doibase 10.1103/PhysRevB.93.115131} {\bibfield  {journal} {\bibinfo
  {journal} {Phys. Rev. B}\ }\textbf {\bibinfo {volume} {93}},\ \bibinfo
  {pages} {115131} (\bibinfo {year} {2016})}\BibitemShut {NoStop}%
\bibitem [{\citenamefont {Shindou}\ \emph {et~al.}(2013)\citenamefont
  {Shindou}, \citenamefont {Matsumoto}, \citenamefont {Murakami},\ and\
  \citenamefont {Ohe}}]{Ohe-ChiralMagnonic}%
  \BibitemOpen
  \bibfield  {author} {\bibinfo {author} {\bibfnamefont {R.}~\bibnamefont
  {Shindou}}, \bibinfo {author} {\bibfnamefont {R.}~\bibnamefont {Matsumoto}},
  \bibinfo {author} {\bibfnamefont {S.}~\bibnamefont {Murakami}}, \ and\
  \bibinfo {author} {\bibfnamefont {J.-i.}\ \bibnamefont {Ohe}},\ }\href
  {\doibase 10.1103/PhysRevB.87.174427} {\bibfield  {journal} {\bibinfo
  {journal} {Phys. Rev. B}\ }\textbf {\bibinfo {volume} {87}},\ \bibinfo
  {pages} {174427} (\bibinfo {year} {2013})}\BibitemShut {NoStop}%
\bibitem [{\citenamefont {Peano}\ \emph {et~al.}(2016)\citenamefont {Peano},
  \citenamefont {Houde}, \citenamefont {Brendel}, \citenamefont {Marquardt},\
  and\ \citenamefont {Clerk}}]{Peano2016}%
  \BibitemOpen
  \bibfield  {author} {\bibinfo {author} {\bibfnamefont {V.}~\bibnamefont
  {Peano}}, \bibinfo {author} {\bibfnamefont {M.}~\bibnamefont {Houde}},
  \bibinfo {author} {\bibfnamefont {C.}~\bibnamefont {Brendel}}, \bibinfo
  {author} {\bibfnamefont {F.}~\bibnamefont {Marquardt}}, \ and\ \bibinfo
  {author} {\bibfnamefont {A.~A.}\ \bibnamefont {Clerk}},\ }\href {\doibase
  10.1038/ncomms10779} {\bibfield  {journal} {\bibinfo  {journal} {Nature
  Communications}\ }\textbf {\bibinfo {volume} {7}},\ \bibinfo {pages} {10779}
  (\bibinfo {year} {2016})}\BibitemShut {NoStop}%
\bibitem [{\citenamefont {Engelhardt}\ and\ \citenamefont
  {Brandes}(2015)}]{Brandes-Bog}%
  \BibitemOpen
  \bibfield  {author} {\bibinfo {author} {\bibfnamefont {G.}~\bibnamefont
  {Engelhardt}}\ and\ \bibinfo {author} {\bibfnamefont {T.}~\bibnamefont
  {Brandes}},\ }\href {\doibase 10.1103/PhysRevA.91.053621} {\bibfield
  {journal} {\bibinfo  {journal} {Phys. Rev. A}\ }\textbf {\bibinfo {volume}
  {91}},\ \bibinfo {pages} {053621} (\bibinfo {year} {2015})}\BibitemShut
  {NoStop}%
\bibitem [{\citenamefont {Ozawa}\ \emph {et~al.}(2019)\citenamefont {Ozawa},
  \citenamefont {Price}, \citenamefont {Amo}, \citenamefont {Goldman},
  \citenamefont {Hafezi}, \citenamefont {Lu}, \citenamefont {Rechtsman},
  \citenamefont {Schuster}, \citenamefont {Simon}, \citenamefont {Zilberberg},\
  and\ \citenamefont {Carusotto}}]{top-phot}%
  \BibitemOpen
  \bibfield  {author} {\bibinfo {author} {\bibfnamefont {T.}~\bibnamefont
  {Ozawa}}, \bibinfo {author} {\bibfnamefont {H.~M.}\ \bibnamefont {Price}},
  \bibinfo {author} {\bibfnamefont {A.}~\bibnamefont {Amo}}, \bibinfo {author}
  {\bibfnamefont {N.}~\bibnamefont {Goldman}}, \bibinfo {author} {\bibfnamefont
  {M.}~\bibnamefont {Hafezi}}, \bibinfo {author} {\bibfnamefont
  {L.}~\bibnamefont {Lu}}, \bibinfo {author} {\bibfnamefont {M.~C.}\
  \bibnamefont {Rechtsman}}, \bibinfo {author} {\bibfnamefont {D.}~\bibnamefont
  {Schuster}}, \bibinfo {author} {\bibfnamefont {J.}~\bibnamefont {Simon}},
  \bibinfo {author} {\bibfnamefont {O.}~\bibnamefont {Zilberberg}}, \ and\
  \bibinfo {author} {\bibfnamefont {I.}~\bibnamefont {Carusotto}},\ }\href
  {\doibase 10.1103/RevModPhys.91.015006} {\bibfield  {journal} {\bibinfo
  {journal} {Rev. Mod. Phys.}\ }\textbf {\bibinfo {volume} {91}},\ \bibinfo
  {pages} {015006} (\bibinfo {year} {2019})}\BibitemShut {NoStop}%
\bibitem [{\citenamefont {Hurst}\ and\ \citenamefont
  {Spielman}(2019)}]{SpielmanFeedback1}%
  \BibitemOpen
  \bibfield  {author} {\bibinfo {author} {\bibfnamefont {H.~M.}\ \bibnamefont
  {Hurst}}\ and\ \bibinfo {author} {\bibfnamefont {I.~B.}\ \bibnamefont
  {Spielman}},\ }\href {\doibase 10.1103/PhysRevA.99.053612} {\bibfield
  {journal} {\bibinfo  {journal} {Phys. Rev. A}\ }\textbf {\bibinfo {volume}
  {99}},\ \bibinfo {pages} {053612} (\bibinfo {year} {2019})}\BibitemShut
  {NoStop}%
\bibitem [{\citenamefont {Hurst}\ \emph {et~al.}(2020)\citenamefont {Hurst},
  \citenamefont {Guo},\ and\ \citenamefont {Spielman}}]{SpielmanFeedback2}%
  \BibitemOpen
  \bibfield  {author} {\bibinfo {author} {\bibfnamefont {H.~M.}\ \bibnamefont
  {Hurst}}, \bibinfo {author} {\bibfnamefont {S.}~\bibnamefont {Guo}}, \ and\
  \bibinfo {author} {\bibfnamefont {I.~B.}\ \bibnamefont {Spielman}},\ }\href
  {\doibase 10.1103/PhysRevResearch.2.043325} {\bibfield  {journal} {\bibinfo
  {journal} {Phys. Rev. Research}\ }\textbf {\bibinfo {volume} {2}},\ \bibinfo
  {pages} {043325} (\bibinfo {year} {2020})}\BibitemShut {NoStop}%
\bibitem [{\citenamefont {Mu\~noz Arias}\ \emph
  {et~al.}(2020{\natexlab{a}})\citenamefont {Mu\~noz Arias}, \citenamefont
  {Poggi}, \citenamefont {Jessen},\ and\ \citenamefont
  {Deutsch}}]{DeutschFeedback1}%
  \BibitemOpen
  \bibfield  {author} {\bibinfo {author} {\bibfnamefont {M.~H.}\ \bibnamefont
  {Mu\~noz Arias}}, \bibinfo {author} {\bibfnamefont {P.~M.}\ \bibnamefont
  {Poggi}}, \bibinfo {author} {\bibfnamefont {P.~S.}\ \bibnamefont {Jessen}}, \
  and\ \bibinfo {author} {\bibfnamefont {I.~H.}\ \bibnamefont {Deutsch}},\
  }\href {\doibase 10.1103/PhysRevLett.124.110503} {\bibfield  {journal}
  {\bibinfo  {journal} {Phys. Rev. Lett.}\ }\textbf {\bibinfo {volume} {124}},\
  \bibinfo {pages} {110503} (\bibinfo {year} {2020}{\natexlab{a}})}\BibitemShut
  {NoStop}%
\bibitem [{\citenamefont {Mu\~noz Arias}\ \emph
  {et~al.}(2020{\natexlab{b}})\citenamefont {Mu\~noz Arias}, \citenamefont
  {Deutsch}, \citenamefont {Jessen},\ and\ \citenamefont
  {Poggi}}]{DeutschFeedback2}%
  \BibitemOpen
  \bibfield  {author} {\bibinfo {author} {\bibfnamefont {M.~H.}\ \bibnamefont
  {Mu\~noz Arias}}, \bibinfo {author} {\bibfnamefont {I.~H.}\ \bibnamefont
  {Deutsch}}, \bibinfo {author} {\bibfnamefont {P.~S.}\ \bibnamefont {Jessen}},
  \ and\ \bibinfo {author} {\bibfnamefont {P.~M.}\ \bibnamefont {Poggi}},\
  }\href {\doibase 10.1103/PhysRevA.102.022610} {\bibfield  {journal} {\bibinfo
   {journal} {Phys. Rev. A}\ }\textbf {\bibinfo {volume} {102}},\ \bibinfo
  {pages} {022610} (\bibinfo {year} {2020}{\natexlab{b}})}\BibitemShut
  {NoStop}%
\bibitem [{\citenamefont {Ozawa}\ and\ \citenamefont
  {Price}(2019)}]{Review-OzawaPrice}%
  \BibitemOpen
  \bibfield  {author} {\bibinfo {author} {\bibfnamefont {T.}~\bibnamefont
  {Ozawa}}\ and\ \bibinfo {author} {\bibfnamefont {H.~M.}\ \bibnamefont
  {Price}},\ }\href {\doibase 10.1038/s42254-019-0045-3} {\bibfield  {journal}
  {\bibinfo  {journal} {Nat. Rev. Phys.}\ }\textbf {\bibinfo {volume} {1}},\
  \bibinfo {pages} {349} (\bibinfo {year} {2019})}\BibitemShut {NoStop}%
\end{thebibliography}%

\end{document}